\documentclass[12pt]{article}
\pdfoutput=1

\usepackage{draft} 
\usepackage[weather]{ifsym}

\usepackage{hyperref}
\usepackage{graphicx,color,subfig}
\usepackage{cite}
\usepackage{mciteplus}
\usepackage{skak}
\usepackage{empheq}
\usepackage{tikz}
\usepackage{bbm}

\DeclareFontFamily{OT1}{pzc}{}
\DeclareFontShape{OT1}{pzc}{m}{it}{<-> s * [1.10] pzcmi7t}{}
\DeclareMathAlphabet{\mathpzc}{OT1}{pzc}{m}{it}

\usetikzlibrary{calc}
\usetikzlibrary{snakes}
\usetikzlibrary{arrows.meta}
\usetikzlibrary{decorations.pathmorphing}
\usetikzlibrary{decorations.markings}
\usetikzlibrary{bending}
\tikzset{snake it/.style={decorate, decoration=snake}}
\usetikzlibrary{shapes.misc}
\tikzset{cross/.style={cross out, draw=black, minimum size=2*(#1-\pgflinewidth), inner sep=0pt, outer sep=0pt},
cross/.default={1pt}}

\usepackage[T1]{fontenc}
\usepackage{esint}
\usepackage{lmodern}

\newcommand{\vv}[1]{\left\langle #1 \right\rangle}
\newcommand{\un}[1]{\underline{#1}}

\def\be#1\ee{\begin{align}#1\end{align}}

\newcommand{\bz}{\bar{z}}

\definecolor{dark green}{rgb}{0.7,1,0.64}

\usepackage{listings}
\usepackage{xcolor}

\definecolor{codegreen}{rgb}{0,0.6,0}
\definecolor{codegray}{rgb}{0.5,0.5,0.5}
\definecolor{codepurple}{rgb}{0.58,0,0.82}
\definecolor{backcolour}{rgb}{0.95,0.95,0.92}

\lstdefinestyle{myStyle}{
    belowcaptionskip=1\baselineskip,
    breaklines=true,
    frame=none,
    numbers=none, 
    basicstyle=\footnotesize\ttfamily,
    keywordstyle=\bfseries\color{green!40!black},
    commentstyle=\itshape\color{purple!40!black},
    identifierstyle=\color{blue},
    backgroundcolor=\color{gray!10!white},
    tabsize=2,
}

\usepackage{array}

\begin{document}

\unitlength = .8mm

\begin{titlepage}

\begin{center}

\hfill \\
\hfill \\
\vskip 1cm

\title{Rolling tachyon and the Phase Space of Open String Field Theory}

\author{Minjae Cho$^{\text{\Lightning}}$, Ben Mazel${}^{\text{\Snow}}$, Xi Yin${}^{\text{\Snow}}$}

\address{
$^{\text{\Lightning}}$Princeton Center for Theoretical Science, Princeton University, \\ Princeton, NJ 08544,
USA
\\
${}^{\text{\Snow}}$Jefferson Physical Laboratory, Harvard University, \\
Cambridge, MA 02138 USA
}

\email{minjae@princeton.edu, bmazel@g.harvard.edu, xiyin@fas.harvard.edu}

\end{center}

\abstract{We construct the symplectic form on the covariant phase space of the open string field theory on a ZZ-brane in $c=1$ string theory, and determine the energy of the rolling tachyon solution, confirming Sen's earlier proposal based on boundary conformal field theory and closed string considerations.
}

\vfill

\end{titlepage}

\eject

\begingroup
\hypersetup{linkcolor=black}

\tableofcontents

\endgroup

\section{Introduction}

The decay process of an unstable D-brane in string theory, known as the rolling tachyon \cite{Sen:2002nu}, is one of the simplest time-dependent backgrounds of string theory and serves as a basic example of open/closed duality. It admits an open string description, either as a time-dependent solution to the open string field theory (OSFT) on the D-brane, or as a boundary conformal field theory (BCFT) on the string worldsheet. Alternatively, it is also expected to admit a dual closed string description that amounts to the closed string radiation resulting from the D-brane decay. 

In the literature, the rolling tachyon has been primarily investigated from the BCFT perspective  \cite{Sen:2002nu,Sen:2002in,Lambert:2003zr,Sen:2004zm,Sen:2004cq,Sen:2004yv,Sen:2004nf}. In bosonic string theory, starting from an unstable D-brane, the rolling tachyon BCFT is obtained through the boundary deformation
\ie\label{deltas}
\Delta S = \widetilde\lambda \int_{\partial\Sigma} \cosh (X^0-u),
\fe
where $X^0$ is a timelike free boson, and $\cosh (X^0-u)$ is a marginal boundary vertex operator defined by the usual boundary normal ordering. Despite its apparent simplicity, the BCFT formulation of the rolling tachyon is subject to a number of subtleties that are ubiquitous to worldsheet string theory in time-dependent backgrounds. First of all, the BCFT is a priori constructed with a Euclidean target space, namely starting from a noncompact free boson $X$ parameterizing the imaginary time coordinate, and then deforming by the marginal boundary coupling $\widetilde\lambda \cos X$. A suitable analytic continuation $X^0-u=-i X$ is then performed at the level of CFT correlators, that in particular involves a contour choice in the target space of $X^0$. A first-principle formulation of physical observables, as well as the interpretation of string amplitudes in this setting, remain to be clarified. 

In this paper, we revisit the rolling tachyon in the OSFT formalism, which is technically more complex but offers a number of conceptual clarifications. To begin with, the rolling tachyon is constructed as a (time-dependent) solution $\Psi$ to the open string field equations. Here the open string field $\Psi$ takes value in the boundary Hilbert space of the undeformed BCFT of the original D-brane. There are several known strategies for constructing such solutions \cite{Kiermaier:2007vu, Erler:2014eqa, Erler:2019fye}. For our purpose, it will suffice to construct $\Psi$ as a perturbative expansion
\ie\label{pertSolone}
\Psi = \sum_{n=1}^\infty \lambda^n \Psi_n,
\fe
where $\lambda$ is related but not equal to $\widetilde\lambda$ in (\ref{deltas}). Both the BCFT and the OSFT solution of the rolling tachyon will be reviewed in section \ref{sec:solution}.

We will focus on the OSFT of a single ZZ-brane in $c=1$ string theory \cite{Zamolodchikov:2001ah, McGreevy:2003kb, Klebanov:2003km}, which is given by Witten's cubic bosonic OSFT \cite{Witten:1985cc} where the matter BCFT consists of a free boson $X^0$ and the vacuum module of a $c=25$ boundary Virasoro algebra. In this setting, the closed string background is consistent at the quantum level, and one may anticipate the OSFT to make sense at the quantum level non-perturbatively, even though the consideration of this paper is restricted to the classical OSFT. In the dual matrix quantum mechanics of $c=1$ string theory \cite{Klebanov:1991qa, Ginsparg:1993is}, the ZZ-brane with rolling tachyon is expected to be described by a single eigenvalue/fermion bouncing off the potential and hovering above the fermi sea \cite{McGreevy:2003kb, Klebanov:2003km}. In this case, one expects the rolling tachyon solutions to cover (at least a domain of) the (covariant) phase space of the OSFT, which is 2-dimensional (parameterized by $\lambda$ and $u$).

A key step in constructing the phase space of the OSFT, as parameterized by its (gauge-inequivalent) solutions, is to determine the symplectic form on the phase space. We adopt a slightly modified version of Witten's proposal \cite{Witten:1986qs}
\ie\label{simpform}
\Omega = {1\over 2g_o^2} \int (\mathbb{P}[Q_B, \Theta(X^0)]) (\delta\Psi*\delta\Psi),
\fe
where $\Theta(X^0)$ is the Heaviside step function in $X^0$ evaluated at the midpoint of the open string, and $\mathbb{P}$ stands for a projector of a local operator onto its weight $(0,0)$ component (see Appendix \ref{sec:osftconvention} for OSFT conventions). We will see in section \ref{sec:symplectic} that such a symplectic form is gauge invariant, is invariant under time translation, and moreover evades a potential singularity that would be encountered in the proposal of \cite{Witten:1986qs} (as was pointed out in \cite{Erler:2004hv}).

We will then evaluate (\ref{simpform}) on the space of rolling tachyon open string field solutions. The result can be expressed in the form
\ie\label{sympEu}
\Omega =  \delta u\delta E(\lambda),
\fe
where $E(\lambda)$ is the energy of the rolling tachyon. As the main result of this paper, in section \ref{sec:energyanalytic} we will give an analytic argument that $E(\lambda)$ agrees with a proposal of Sen \cite{Sen:2002nu,Sen:2004yv} based on the BCFT and the conserved charges for the open string fields induced by the rigid gauge transformations of closed string fields in open-closed string field theory (OCSFT). Our analytic result is further supported by a highly nontrivial level truncation computation at order $\lambda^4$ in section \ref{sec:energyperturb}.

More broadly, we anticipate the construction of the symplectic form and the Hamiltonian on the covariant phase space to serve as a step toward defining the quantum OSFT, both at the level of time-dependent perturbation theory (e.g. around the rolling tachyon background), and at the non-perturbative level. Some future perspectives are discussed in section \ref{sec:discuss}.

\section{Rolling tachyon}
\label{sec:solution}

In this section, we review the construction of the rolling tachyon solution in classical OSFT and the corresponding BCFT description, following \cite{Sen:2004nf}. Witten's cubic form of the bosonic OSFT is defined through the action
\cite{Witten:1985cc} 
\ie\label{OSFTaction}
S[\Psi] = -{1\over g_o^2}\int\left({1\over2} \Psi*Q_B\Psi+{1\over3}\Psi*\Psi*\Psi \right),
\fe
where the (classical) open string field $\Psi$ is a ghost number 1 state of the worldsheet BCFT, $Q_B$ is the BRST operator. The star product $*$ and the convolution $\int$ on the open string fields are defined in Appendix \ref{sec:osftconvention}. Variation of (\ref{OSFTaction}) with respect to $\Psi$ gives the classical equation of motion
\ie\label{OSFTEOM}
Q_B\Psi+\Psi*\Psi = 0.
\fe
Specifically, we will consider the OSFT on a single type $(1,1)$ ZZ-brane in $c=1$ string theory \cite{McGreevy:2003kb, Klebanov:2003km, Sen:2004nf}. The worldsheet (bulk) CFT consists of a timelike free boson $X^0$, the $c=25$ Liouville CFT, and the $c=-26$ $bc$-ghost system. The ZZ-brane is described by a BCFT on the worldsheet that is Neumann with respect to $X^0$, and the type $(1,1)$ ZZ boundary condition in the Liouville sector \cite{Zamolodchikov:2001ah}. The latter may be characterized by its boundary state, or equivalently the disc 1-point functions (following the convention of \cite{Balthazar:2019rnh})
\ie
\langle V_P \rangle^{D^2}_{ZZ}=2^{5\over4}\sqrt{\pi}\text{sinh}(2\pi P),
\fe 
where $V_P$ is a bulk primary of the Liouville CFT that carries Liouville momentum $P$, and corresponding holomorphic and anti-holomorphic weights $h=\widetilde h = 1+P^2$. The boundary operator spectrum of the ZZ-boundary Liouville theory is extremely simple: namely, it consists of only the identity operator and its Virasoro descendants. This identity primary of the boundary Liouville theory gives rise to the open string tachyon on the ZZ-brane, whose corresponding vertex operator takes the form $c e^{i\omega X^0} (\otimes 1)$.

The rolling tachyon on the ZZ-brane is represented by a time-dependent solution $\Psi$ to the OSFT equation (\ref{OSFTEOM}), which may be constructed perturbatively in the form (\ref{pertSolone}) where the $\Psi_n$'s are solved successively at each order in $\lambda$,
\ie{}
\lambda^1:&~~Q_B\Psi_1=0,
\\
\lambda^2:&~~Q_B\Psi_2+\Psi_1*\Psi_1=0,
\\
\lambda^3:&~~Q_B\Psi_3+\Psi_1*\Psi_2+\Psi_2*\Psi_1=0,
\\
...
\fe
At the linearized order, $\Psi_1$ takes the form of an on-shell tachyon vertex operator, which we will take to be\footnote{Here and henceforth the boundary normal ordering is understood in all expressions for boundary vertex operators in the $X^0$ CFT. We have also adopted the convention $\A'=1$.}
\ie\label{linearterm}
\Psi_1&=c\cosh(X^0-u).
\fe
Such a solution represents a tachyon field that starts rolling at time $x^0=u$ with magnitude $\sim \lambda$. 
The higher order terms of the open string field are solved in Siegel gauge via
\ie\label{pertSol}
\Psi_n&=-{b_0\over L_0}\sum_{k=1}^{n-1}\Psi_k*\Psi_{n-k},~~~n\geq 2.
\fe
In this procedure, a priori one may encounter obstructions when the star product $\Psi_k*\Psi_{n-k}$ contains weight zero states, on which ${1\over L_0}$ is ill-defined. However, the exact marginality of $\text{cosh}(X^0-u)$ (viewed as a BCFT deforrmation) guarantees that such obstructions are absent to all orders in $\lambda$. Indeed, the perturbative solution (\ref{linearterm}), (\ref{pertSol}) is valid to all orders in $\lambda$, even though the radius of convergence with respect to $\lambda$ is believed to be finite (see \cite{Kudrna:2019xnw}, chapter 6).

The rolling tachyon may alternatively be described as an exactly marginal deformation of the worldsheet BCFT of the form (\ref{deltas}). As usual in conformal perturbation theory, the precise definition of (\ref{deltas}) as a deformation depends on a choice of regularization scheme. We will adopt the scheme in which a correlator on the upper half plane (UHP) subject to the deformed boundary condition is computed as
\ie
\langle \cdots\rangle_{\rm deformed}^{\rm UHP} = \sum_{n=0}^\infty {\tilde\lambda^n\over n!} \bigg\langle \cdots \prod_{k=1}^{n} \int_{W_k} {dt_k\over2\pi}\text{cosh}(X^0(t_k)-u)\bigg\rangle^{\rm doubling}_{\rm undeformed},
\fe
where $\langle ... \rangle^{\rm doubling}$ stands for the correlator computed using the doubling trick for $X^0$ CFT (i.e. replacing the free boson on the UHP with a chiral boson on the entire complex plane), and the integration contours $W_k$ of the boundary operators are taken to be $\mathbb{R}+i\E_k$ in the complex plane obtained by the doubling trick, for distinct small real parameters $\epsilon_1,\cdots, \epsilon_n$.  The exact boundary state in the $X^0$ CFT is known \cite{Callan:1994ub,Recknagel:1998ih}, and takes the form\footnote{This boundary state was determined in \cite{Callan:1994ub,Recknagel:1998ih} for the ${\tilde\lambda}\text{cos}X$ deformation of the Neumann boundary condition of a spacelike free boson $X$, and then analytically continued to that of the timelike free boson $X^0$ in \cite{Sen:2002nu}.}
\ie\label{bxzero}
|{\cal B}\rangle_{X^0}=\bigg[f(X^0)+\left(\text{cos}(2\pi{\tilde\lambda})+1-f(X^0)\right)\A_{-1}{\bar\A}_{-1}+ \cdots \bigg] |0\rangle,
\fe
where $f(X^0)={1\over1+e^{X^0}\text{sin}(\pi{\tilde\lambda})}+{1\over1+e^{-X^0}\text{sin}(\pi{\tilde\lambda})}-1$ and $\cdots$ stands for terms at higher oscillator levels. Note that $\tilde\lambda$ is not the same as the parameter $\lambda$ appearing in (\ref{OSFTEOM}), even though they agree to first order; their precise relation ${\tilde\lambda}={\tilde\lambda}(\lambda)$ will be discussed in section \ref{sec:comparisonToSen}. In the full BCFT of the string worldsheet, the ZZ boundary condition of the $c=25$ Liouville theory as well as that of the $bc$ ghosts are undeformed; the full resulting boundary state will be denoted by $|{\cal B}_{\tilde\lambda, u}\rangle$. 

The background independence of OSFT \cite{Sen:1990hh} is such that every exactly marginal deformation of the BCFT can be equivalently represented as a solution to the original OSFT equation defined in the undeformed BCFT. In the original OSFT on the ZZ-brane, an exact solution $\Psi(\widetilde\lambda, u)$ to the equation of motion (\ref{OSFTEOM}) that is physically equivalent to the deformed BCFT has been constructed in \cite{Kiermaier:2007vu}. Furthermore, the on-shell components of the boundary state $|{\cal B}_{\tilde\lambda, u}\rangle$ can be directly related to the string field solution through the so-called Ellwood invariant $W$ \cite{Ellwood:2008jh}, via
\ie
\label{ellwoodConj}
{i\over4\pi}\left(\langle{\cal B}_{\tilde\lambda, u}|(c_0-{\tilde c}_0)|V\rangle-\langle{\cal B}_{\tilde\lambda=0}|(c_0-{\tilde c}_0)|V\rangle\right) \equiv W(\Psi(\tilde\lambda, u),V) = \langle \bI|V(i)|\Psi(\tilde\lambda, u)\rangle.
\fe
Here $V$ is an arbitrary bulk vertex operator which is $Q_B$-closed, and  $\bI$ is the ``identity string field" defined as the identity element with respect to the $*$-algebra \cite{Rastelli:2000iu}.\footnote{Precisely speaking, \cite{Ellwood:2008jh} showed that (\ref{ellwoodConj}) holds true assuming that $V$ takes the form $c\tilde{c}V_m$, where $V_m$ is a matter primary of weight (1,1). Nonetheless, (\ref{ellwoodConj}) is expected to be true for any $Q_B$-closed $V$, based on the background independence of string field theory \cite{Sen:1990hh}. When $V$ is not a conformal primary of weight (0,0), the RHS of  (\ref{ellwoodConj}) may be singular in the cubic OSFT, but a more general open-closed string vertices will provide a suitably regularized result.} The matrix element on the RHS of (\ref{ellwoodConj}) is understood in radial quantization on the UHP, with the string field $\Psi(\widetilde\lambda, u)$ inserted at the origin, the identity string field $\bI$ inserted at infinity, and the vertex operator $V$ inserted at $i$. An equivalent expression for the Ellwood invariant is
\ie\label{ellwoodUHP}
W(\Psi(\tilde\lambda, u),V)=\langle V(i)f\circ\Psi(\tilde\lambda, u)(w=0) \rangle^{\text{UHP}},
\fe
where $f\circ\Psi(\tilde\lambda, u)$ is the conformal transformation of $\Psi(\tilde\lambda, u)$ under the map 
\ie\label{zfwmap}
z=f(w)\equiv{2w\over1-w^2}
\fe
which takes the half unit disc to the UHP.

As was explained in \cite{Sen:2004yv}, the boundary state $|{\cal B}_{{\tilde\lambda},u}\rangle$ captures the spacetime conserved charges associated with the solution $\Psi({\tilde\lambda},u)$, as can be seen from the rigid gauge transformations of the closed string fields which induce symmetry transformations of the open string fields in open+closed SFT \cite{Zwiebach:1997fe,Sen:2004yv}. For instance, consider a gauge variation of the closed string field $\Phi$ of the form $\delta\Phi=Q_B\Lambda(\omega)$, with $\Lambda(\omega)=\left(c\partial X^0-{\tilde c}{\bar\partial}X^0\right)e^{i\omega X^0}$. The ``rigid'' part $\Lambda(0)$ is $Q_B$-closed and represents an isometry of the closed string background $\Phi=0$, which in this case amounts to time translation symmetry. It further induces time translation on the open string fields, through the closed string tadpole terms in open+closed SFT action which are linear in the boundary state $|{\cal B}_{{\tilde\lambda},u}\rangle$. The corresponding conserved charge $E$, which has the interpretation of energy, is then proportional to
\ie\label{SenCharge}
\int {dp\over2\pi}e^{px^0}\langle{\cal B}_{{\tilde\lambda},u}|(c_0-{\tilde c}_0)|\phi(p)\rangle,
\fe
where $x^0$ is the zero mode of $X^0$, and $\phi(p)$ is defined by $Q_B\Lambda(p)=p\phi(p)$ and thus given by
\ie\label{phiterms}
|\phi(p)\rangle\sim\left[2c_1{\tilde c}_1\alpha_{-1}{\tilde\A}_{-1}+c_{-1}c_1-{\tilde c}_{-1}{\tilde c}_1+\cdots\right]|p\rangle,
\fe
where $\cdots$ are terms linear in $p$. The integration over $p$ in (\ref{SenCharge}) actually localizes to $p=0$ and the final result is
\ie\label{SenEnergy}
E(\lambda({\tilde\lambda}))={1\over 4\pi^2g_o^2}(1+\text{cos}(2\pi\tilde\lambda)).
\fe
In particular, the energy $E(\lambda({\tilde\lambda}))$ is determined by the coefficients of the zero momentum dilaton vertex operator 
\ie\label{vddef}
V_D=c\tilde c\partial X^0\bar\partial X^0,
\fe
and that of the zero momentum ghost-dilaton vertex operator $c\partial^2c-{\tilde c}{\bar\partial}^2{\tilde c}$, in the boundary state $|{\cal B}_{{\tilde\lambda},u}\rangle$. The $\tilde\lambda$ dependence of $E(\lambda({\tilde\lambda}))$ comes purely from $V_D$, whose Ellwood invariant is given by \footnote{For generic OSFT backgrounds whose worldsheet theory includes $X^0$-CFT, this particular Ellwood invaraint, when $\Psi(\tilde\lambda,u)$ is replaced by any static solution $\Psi_s$ of the OSFT EOM (\ref{OSFTEOM}) satisfying some regularity conditions, was shown to compute the energy of the solution $\Psi_s$ in the sense that it agrees with the OSFT action (\ref{OSFTaction}) evaluated on $\Psi=\Psi_s$ \cite{Baba:2012cs}.}
\ie\label{ellwoodDilaton}
W(\Psi(\tilde\lambda,u),V_D)={V_{X^0}\over4\pi i}\text{cos}(2\pi\tilde\lambda)
\fe
where $V_{X^0}$ is the volume of $X^0$. In section \ref{sec:energyanalytic}, we will obtain the energy expression (\ref{SenEnergy}) by showing how $W(\Psi(\tilde\lambda,u),V_D)$ arises from the symplectic form of the classical OSFT.

\section{A symplectic form for OSFT}
\label{sec:symplectic}

A symplectic form on the phase space of the bosonic OSFT was proposed by Witten \cite{Witten:1986qs}. In this section, we review the construction of \cite{Witten:1986qs} and propose a slightly modified regularized version of this symplectic form. 

We begin by briefly reviewing the covariant phase space formalism.\footnote{See e.g. \cite{Harlow:2019yfa} for an introduction to the subject.} For our purpose, it suffices to restrict to the case where the spatial manifold has no boundary. Given a Lagrangian field theory with the action $S=\int_M L$, where $M$ is the spacetime manifold and $L$ is a function of the fields $\phi^a$ and their derivatives, the classical pre-phase space is the space of solutions to the equations of motions $e_a=0$ obtained by varying the Lagrangian
\ie\label{ldu}
\delta L = e_a \D\phi^a+dU.
\fe
Note that the total derivative term $dU$ is linear in the variation of $\phi^a$. $U$ is known as the pre-symplectic potential. The pre-symplectic form is defined as 
\ie
\widetilde\Omega = \int_\Sigma \delta U, 
\fe
where $\Sigma$ is a Cauchy slice and $\delta$, which originally denoted the field variation, is now interpreted as the exterior derivative on the pre-phase space.\footnote{$\delta$ will henceforth be regarded as a Grassmann-odd operation. } In particular, $\widetilde\Omega$ is a closed 2-form on the pre-phase space. 

In a gauge theory, the phase space is obtained as the quotient of the pre-phase space of solutions by the group of gauge transformations, and is equipped with the symplectic form $\Omega$ induced from the pre-symplectic form $\widetilde\Omega$. The contraction of $\Omega$ with a Hamiltonian vector field produces the differential of the corresponding conserved charge. In the case of the rolling tachyon solution $\Psi(\tilde\lambda, u)$, the Hamiltonian vector field ${\D\over\D u}$ generates time translation, and the corresponding energy $E$ is related by $\Omega\left({\D\over\D u},\cdot\right)=\D E$.

\subsection{Witten's construction of the symplectic form}

Following the general covariant phase space prescription outlined above, we consider the variation of the OSFT action (\ref{OSFTaction})
\ie
\D S[\Psi] = -{1\over g_o^2}\int\left[ (Q_B\Psi+\Psi*\Psi)*\D\Psi-{1\over2}Q_B(\Psi*\D\Psi) \right].
\fe
Indeed, as in (\ref{ldu}), the total derivative term ${1\over 2g_o^2}\int Q_B(\Psi*\D\Psi)$ gives rise to the pre-symplectic potential $U$ which will then be used to construct the pre-symplectic form $\widetilde\Omega$.

Recall that $Q_B$ is a second order derivative operator with respect to the (space-)time coordinate (as the BRST current takes the form $j_B\sim c\partial X^0\partial X^0+\cdots$), while the pre-symplectic potential $U$ appears in  (\ref{ldu}) through its first order derivative. Thus in extracting $U$ from the expression $\delta L\sim Q_B(\Psi*\D\Psi)$, one needs to ``strip off'' one derivative from $Q_B$. Witten's prescription \cite{Witten:1986qs} amounts to constructing $U$ by replacing $Q_B$ in ${1\over 2g_o^2}\int Q_B(\Psi*\D\Psi)$ with the commutator of $Q_B$ with the Heaviside step function $\Theta(\Sigma)$ with respect to the open string midpoint. Namely, $\Theta(\Sigma)$ is defined
to be $1$ if the open string midpoint is to the future of the Cauchy slice $\Sigma$ and $0$ otherwise. The resulting pre-symplectic form is
\ie\label{OSFTsympW}
\Omega_W=\D\left({1\over2g_o^2}\int [Q_B,\Theta(\Sigma)](\Psi*\D\Psi) \right)={1\over2g_o^2}\int [Q_B,\Theta(\Sigma)](\D\Psi*\D\Psi),
\fe
which may equivalently be written in the form of a correlator on the UHP
\ie\label{OSFTsympWUHP}
\Omega_W={1\over2g_o^2}\bigg\langle [Q_B,\Theta(\Sigma)](i) ~f_1\circ\D\Psi(w_1=0) ~f_2\circ\D\Psi(w_2=0)\bigg\rangle^{\rm UHP}.
\fe
Here $f\circ \delta\Psi$ stands for the conformal transformation of the string field $\delta\Psi$ with respect to the map $z=f(w)$, with the choice
\ie
f_1(w)={1+w\over1-w},~~~~f_2(w)=-{1-w\over1+w}.
\fe
In \cite{Witten:1986qs}, $\Omega_W$ was argued to be independent of the choice of $\Sigma$ and gauge invariant. However, the definition of $\Omega_W$ via the star product implicitly involves a conformal transformation that acts singularly on the operator $[Q_B,\Theta(\Sigma)]$ which is inserted at the midpoint of the string. For generic choices of $\Sigma$, such as constant time slicing $X^0=const$, $[Q_B,\Theta(\Sigma)]$ would not be a primary of weight $(0,0)$, whose singular conformal transformation results in $\Omega_W$ being ill-defined.\footnote{In a string background where there is at least one spatial coordinate $Y$ and a lightlike isometry, the lightcone time slicing $X^+=X^0+Y=const$ provides a primary of weight $(0,0)$ at the midpoint since $e^{ik_+X^+}$ is a primary of weight $(0,0)$ for any $k_+$. We thank Ted Erler for bringing this point and the relevant work \cite{Erler:2004hv} to our attention.}

\subsection{A regularized symplectic form for the OSFT}\label{OSFTsympForm}

A simple fix for the singularity in the definition of $\Omega_W$ is to replace it with 
\ie\label{OSFTsymp}
\Omega&={1\over2g_o^2}\int \left({\mathbb P}[Q_B,\Theta(\Sigma)]\right)(\D\Psi*\D\Psi)
\\
&={1\over2g_o^2}\bigg\langle \left({\mathbb P}[Q_B,\Theta(\Sigma)]\right)(i) ~f_1\circ\D\Psi(w_1=0) ~f_2\circ\D\Psi(w_2=0)\bigg\rangle^{\rm UHP},
\fe
where $\mathbb P$ stands for the projector that takes a local operator to its weight $(0,0)$ component.\footnote{More precisely, $\mathbb P$ can be defined by orthogonal projection to the subspace of a definite weight in the matter CFT Hilbert space, when restricted to a given oscillator level with respect to the $bc$ ghost system.} Note that we should be cautious in commuting $\mathbb{P}$ with $Q_B$, as $\mathbb P$ is not well-defined when acting directly on $\Theta(\Sigma)$, e.g. for a constant time slicing $X^0=v$,
\ie
\Theta(\Sigma_{X^0=v})={1\over2\pi i}\int_{-\infty}^{\infty} {1\over p-i\E} e^{ip (X^0-v)}dp.
\fe
Nonetheless, it makes sense to consider the $\mathbb P$-projection of $[Q_B,\Theta(\Sigma_{X^0=v})]$,
\ie\label{defD}
D\equiv {\mathbb P}[Q_B,\Theta(\Sigma_{X^0=v})]&={\mathbb P}{1\over2\pi}\int_{-\infty}^\infty\left(c\partial X^0+{\tilde c}{\bar\partial}X^0+ \cdots\right)e^{ip (X^0-v)}dp
\\
&={1\over V_{X^0}}\int_{-\infty}^{\infty} {\delta(p)}\left(c\partial X^0+{\tilde c}{\bar\partial}X^0+\cdots\right)e^{ip (X^0-v)}dp
\\
&={1\over V_{X^0}}\left(c\partial X^0+{\tilde c}{\bar\partial}X^0\right),
\fe
where the omitted terms in the first and second lines are linear in $p$ and drop out after the projection, and $V_{X^0}$ stands for the time volume which will end up canceling against a similar volume factor from the CFT correlator. 
More generally, $\mathbb P$ is well defined on an operator of the form $f(X^0)$ provided that the function $f$ has compact support. In particular, given two Cauchy slices $\Sigma_1$ and $\Sigma_2$, we can write
\ie
{\mathbb P}[Q_B,\Theta(\Sigma_1)-\Theta(\Sigma_2)]=[Q_B,{\mathbb P}\left(\Theta(\Sigma_1)-\Theta(\Sigma_2)\right)].
\fe

We now proceed to prove several properties of $\Omega$ (\ref{OSFTsymp}), including its gauge invariance. The latter allows for defining the symplectic form simply by restricting the pre-symplectic form to a set of gauge-fixed solutions to the OSFT equation of motion. With this understanding, we will henceforth not distinguish between the notations for the symplectic form and the pre-symplectic form.

\subsubsection{Independence of $\Sigma$}\label{subsubsec:indepSigma}

The a priori definition of $\Omega$ (\ref{OSFTsymp}) necessitates a choice of $\Sigma$. Nevertheless, it proves to be independent of this choice, as follows. The difference between $\Omega$ defined with respect to two slices $\Sigma_1$ and $\Sigma_2$ can be written as
\ie\label{indsigm}
&\int\mathbb{P}[Q_B,\Theta(\Sigma_1)-\Theta(\Sigma_2)](\delta\Psi*\delta\Psi)=\int[Q_B,\mathbb{P}\left(\Theta(\Sigma_1)-\Theta(\Sigma_2)\right)](\delta\Psi*\delta\Psi)
\\
&=\int Q_B\bigg(\left(\mathbb{P}\left(\Theta(\Sigma_1)-\Theta(\Sigma_2)\right)\right)(\delta\Psi*\delta\Psi)\bigg)-\int \mathbb{P}\left(\Theta(\Sigma_1)-\Theta(\Sigma_2)\right)Q_B(\delta\Psi*\delta\Psi).
\fe
The first term on the second line of (\ref{indsigm}) is the integral of a total derivative that vanishes as $\left(\mathbb{P}\left(\Theta(\Sigma_1)-\Theta(\Sigma_2)\right)\right)(\delta\Psi*\delta\Psi)$ has no support at the boundaries of spacetime. The second term on the second line of (\ref{indsigm}) also vanishes, as follows from the equation of motion $Q_B\Psi+\Psi*\Psi=0$ and the cyclicity of the convolution with respect to the star product, in the presence of a bulk insertion at the string midpoint: 
\ie{}
&\int \mathbb{P}\left(\Theta(\Sigma_1)-\Theta(\Sigma_2)\right)Q_B(\delta\Psi*\delta\Psi)= - 2\int \mathbb{P}\left(\Theta(\Sigma_1)-\Theta(\Sigma_2)\right)(\delta (Q_B\Psi)*\delta\Psi)
\\
&= 2\int \mathbb{P}\left(\Theta(\Sigma_1)-\Theta(\Sigma_2)\right)\left(\delta\Psi*\Psi*\delta\Psi - \Psi*\delta\Psi*\delta\Psi \right)=0.
\fe
Note that in these manipulations we have used the Grassmann-oddness of the exterior differential $\delta$ as well as the string field $\Psi$ itself.

\subsubsection{Invariance with respect to the gauge orbit}

The classical OSFT action is invariant under the gauge transformation
\ie
\Psi\mapsto \Psi+Q_B\E+\E*\Psi-\Psi*\E,
\fe
where the gauge parameter $\E$ is a string field of ghost number zero. When $\Psi$ solves the equation of motion and thereby corresponds to a point in the pre-phase space, $\E$ generates the gauge orbit of this solution, and we expect the pre-symplectic form $\Omega$ to be invariant along the gauge orbit. To see this, note that the differential $\delta\Psi$ on the pre-phase space transforms under $\E$ by
\ie
\delta\Psi\mapsto\delta\Psi+\E*\delta\Psi-\delta\Psi*\E,
\fe
and thus 
\ie
\Omega\mapsto\Omega+2\int{\mathbb P}[Q_B,\Theta(\Sigma)]\left(\E*\delta\Psi*\delta\Psi-\delta\Psi*\E*\delta\Psi \right)=\Omega,
\fe
due to the aforementioned cyclicity property.

\subsubsection{Decoupling of null tangent vectors}

At each point $\Psi$ in the pre-phase space, we define 
\ie
Q_\Psi \equiv Q_B+[\Psi,\cdot\}_*
\fe 
where $[\cdot,\cdot\}_*$ stands for the graded commutator with respect to the star product. It follows from the equation of motion that $Q_\Psi$ is nilpotent. Furthermore, tangent vectors of the pre-phase space at $\Psi$ are in correspondence with $Q_\Psi$-closed string fields, whereas $Q_\Psi$-exact string fields amount to gauge redundancies, or ``null tangent vectors''.  

We expect the pre-symplectic form $\Omega$ to vanish upon contraction with a null tangent vector. In other words, if we replace one of $\delta\Psi$'s in $\Omega$ with $Q_\Psi\E = Q_B\E+\Psi*\E-\E*\Psi$, for a ghost number zero string field $\E$, the result should vanish. Indeed,
\ie{}
&\int{\mathbb P}[Q_B,\Theta(\Sigma)](\delta\Psi* Q_\Psi\E )=\int{\mathbb P}[Q_B,\Theta(\Sigma)](\delta (Q_B\Psi)*\E+\delta\Psi*[\Psi,\E\}_*)
\\
&=\int{\mathbb P}[Q_B,\Theta(\Sigma)]((-\delta\Psi*\Psi+\Psi*\delta\Psi)*\E+\delta\Psi*[\Psi,\E\}_*)
=0,
\fe
where we have used $Q_B^2=0$ in the first equality, $Q_B\Psi+\Psi*\Psi=0$ in the second quality, and cyclicity in the third equality.

\bigskip

In summary, we have shown that $\Omega$ as defined in (\ref{OSFTsymp}) is independent of the choice of the Cauchy slice $\Sigma$, is invariant along the gauge orbit of a string field solution, and vanishes upon contraction with a null tangent vector on the pre-phase space. It follows that $\Omega$ induces a well-defined symplectic form on the phase space, parameterized by gauge-equivalence classes of solutions to the OSFT equation of motion.

\section{Perturbative evaluation of the symplectic form}
\label{sec:energyperturb}

In this section, we present an explicit perturbative evaluation of the symplectic form $\Omega$ (\ref{OSFTsymp}), (\ref{defD}) on the phase space of the rolling tachyon on the ZZ-brane, based on level truncation approximation.

Recall from section \ref{sec:solution} that the rolling tachyon solutions of the OSFT on the ZZ-brane are parameterized by $(u, \lambda)$ where $\lambda$ is the natural expansion parameter of the perturbative string field solution, or $(u, \tilde\lambda)$ where $\tilde\lambda$ is the natural deformation parameter in the BCFT description. The corresponding two-dimensional phase space is equipped with a symplectic form $\Omega$ of the form (\ref{sympEu}), where $E(\lambda)$ is the energy of the solution up to a constant shift. Note that $\Omega$ is invariant under shift of $u$.

When evaluated in terms of the perturbative rolling tachyon solution (\ref{pertSol}), $\Omega$ takes the form of a power series expansion in $\lambda$,
\ie
\Omega=\sum_{n=0}^\infty \Omega^{(n)} \equiv \sum_{n=0}^\infty n E_{n} \lambda^{n-1} \delta u\delta\lambda ,
\fe
with the energy expanded as $E(\lambda) = \sum_{n=0}^\infty E_n \lambda^n $.
In particular, the solution at $\lambda=0$ corresponds to the unperturbed ZZ-brane, whose energy is $E_0={1\over2\pi^2g_o^2}$.

\subsection{Leading order in $\lambda$}

We begin with the leading nontrivial order in $\lambda$, namely $\Omega^{(2)}$ or equivalently the energy coefficient $E_2$. The relevant string field solution, to first order in $\lambda$, is given by $\Psi=\lambda ~c~ \text{cosh}(X^0-u)+{\cal O}(\lambda^2)$. Substituting its variation $\delta\Psi=\delta\lambda~c~\text{cosh}(X^0-u)-\delta u~\lambda~c~\text{sinh}(X^0-u)+{\cal O}(\lambda^2)$ into  (\ref{OSFTsymp}) leads to
\ie{}
\Omega^{(2)} ={\lambda\D u\D\lambda\over2g_o^2}\bigg\langle &D(i) ~\bigg(c~\text{sinh}(X^0-u)(z=1) ~c~\text{cosh}(X^0-u)(z=-1)
\\
&-~c~\text{cosh}(X^0-u)(z=1) ~c~\text{sinh}(X^0-u)(z=-1)\bigg)\bigg\rangle^{\rm UHP},
\fe
where $D$ is defined as in (\ref{defD}). Using the following correlators of the ghost and matter CFTs on the UHP,\footnote{We have adopted an overall normalization convention that is compatible with the unitarity of open string amplitudes.}
\ie{}
&\langle c(z_1)c(z_2)c(z_3)\rangle= |(z_1-z_2)(z_1-z_3)(z_2-z_3)|,
\\
&\bigg\langle \prod_n e^{k_n X^0}(y_n)\bigg\rangle=2\pi \D\left(\sum_n k_n\right)\prod_{n>m}|y_n-y_m|^{2k_nk_m},
\fe
we obtain
\ie
\Omega^{(2)} =-{\lambda\D u\D\lambda\over g_o^2}=\delta u\delta\left(-{\lambda^2\over2g_o^2}\right).
\fe
It follows that the energy of the perturbative solution to order $\lambda^2$ is
\ie
E(\lambda)={1\over2\pi^2g_o^2}\left(1-\pi^2\lambda^2+{\cal O}(\lambda^4)\right).
\fe
With the identification $\lambda=\tilde\lambda+{\cal O}(\tilde\lambda^3)$, this result is in agreement with Sen's expression (\ref{SenEnergy}).

\subsection{Subleading order in $\lambda$}
\label{sec:leveltrunc}

The next order term in the $\lambda$-expansion of the symplectic form is $\Omega^{(4)}$,\footnote{Note that $\Omega^{(k)}=0$ for all odd $k$, due to time reversal invariance of the perturbative solution.} which determines the energy coefficient $E_4$. In this subsection we will aim to evaluate the quantity $r \equiv {E_4 \over E_2}$ numerically with the level truncation method, which has been successfully applied in various aspects of SFT in the past \cite{Sen:1999nx,Moeller:2000xv,Moeller:2000jy,Rastelli:2000iu,Kudrna:2012re,Kudrna:2019xnw}. Namely, we will truncate the space of open string fields to a finite dimensional subspace graded by the worldsheet oscillator level, evaluate the symplectic form in this approximation, and numerically extrapolate the result to the limit where the truncation level is taken to infinity.

We begin by specifying a basis of the BCFT Hilbert space as follows. The open string field $\Psi$ lies in the ghost number 1 subspace of boundary operators in the BCFT Hilbert space. For the $(1,1)$ ZZ-brane in question, the BCFT Hilbert space is the tensor product of the $c=25$ identity Virasoro module in the Liouville sector, the Hilbert space of Neumann boundary operators in the $X^0$ free boson sector, and that of the $bc$ ghost system subject to the standard (Neumann type) boundary condition. We will work with the basis states
\ie
\label{generalState}
L^{(25)}_{-\un{K}} \ket{1}_\text{Liouville} \otimes \alpha_{-\un{M}} \ket{f(X^0)}_{X^0} \otimes b_{-\un{N}} c_{-\un{P}} \ket{\downarrow}_{bc},
\fe
where $\un{K}=\{k_1, \cdots, k_{n_K}\}$ is a set of non-negative integers with $k_1 \geq ... \geq k_{n_K}$, and $L^{(25)}_{-\un{K}}$ stands for the product of a sequence of operators $L^{(25)}_{-k_1} \cdots L^{(25)}_{-k_{n_K}}$ (similarly for $ \alpha_{-\un{M}},b_{-\un{N}},$ and $c_{-\un{P}}$). $L^{(25)}_{-k}$ are the $c=25$ Virasoro raising operators in the Liouville sector (note that $L^{(25)}_{-1}$ is not needed as it annihilates the identity boundary primary), and $\A_{-m}, b_{-n}, c_{-p}$ are the usual oscillators in the free boson and the ghost BCFT. The state $|f(X^0)\rangle_{X^0}$ corresponds to the boundary normal ordered operator $f(X^0)$. $\ket{\downarrow}_{bc} = c_1 |1\rangle_{bc}$ is the $bc$ ghost ground state annihilated by $b_m$ for $m\geq0$, and we will impose Siegel gauge condition by omitting $c_0$ from the oscillators appearing in (\ref{generalState}). The {\it level} of such a basis state is defined as 
\ie
|K|+|M|+|N|+|P|, 
\fe
where $|K| \equiv k_1 + \cdots + k_{n_K}$, and similarly for $|M|$, $|N|$, $|P|$.

The star product of a pair of string fields of the form (\ref{generalState}) is conveniently evaluated using a set of Ward identities, known as the ``cubic vertex conservation laws'' in the cubic OSFT \cite{Kudrna:2019xnw}, which replace a raising operator acting on one of the three string fields in a cubic vertex with a sum of lowering operators acting on all three string fields. Iterating the latter procedure, we can reduce any cubic vertex of string fields of the form (\ref{generalState}) to that of (boundary) conformal primaries. Given three primaries $\phi_i$ with conformal weights $h_i$, $i=1,2,3$, their cubic vertex evaluates to simply
 \ie
 \{ \phi_1, \phi_2, \phi_3 \} = C_{123} K^{-(h_1+h_2+h_3)},
 \fe
where $C_{123}$ is the boundary structure constant, and the power of $K = \frac{3\sqrt{3}}{4}$ appears due to the conformal mapping involved in the definition of Witten's cubic vertex.
 
The level truncation approximation to (\ref{pertSol}) proceeds by restricting the string fields to the subspace spanned by the basis states (\ref{generalState}) up to a cutoff level ${\bf L}$, as well as truncating the star products of such string fields up to the same level, with the expectation that observables computed with the level-truncated string fields converge to their exact values in the ${\bf L}\to \infty$ limit. For instance, consider the first step of (\ref{pertSol}), 
\ie
\label{eq:Psi2Def}
\Psi_2 = -\frac{b_0}{L_0} \widetilde{\Psi}_2,~~~~\widetilde{\Psi}_2\equiv \Psi_1 * \Psi_1.
\fe
Let $|v, f\rangle$ be a Siegel gauge ($0 \notin \un{P}$) ghost-number 1 basis state of the form (\ref{generalState}), labeled by $v\equiv (\un{K},\un{M},\un{N},\un{P})$ and the function $f(X^0)$. Its level is denoted $|v| \equiv |K|+|M|+|N|+|P|$. We define a set of dual basis states $\langle v^c, f|$ by the property $\vv{v^c, f|c_0|v',f'} = \delta_{v,v'} (f,f')$, where the pairing $(f,f')$ is given by integration with respect to $X^0$, and $\ket{v^c,f}$ the BPZ conjugate of $\bra{v^c,f}$. $\widetilde{\Psi}_2$ is a ghost-number 2 string field of the form
\ie
 \ket{\widetilde{\Psi}_2} = \sum_{v,\, f,\, |v| \leq {\bf L}} A_{2,v,f} c_0 \ket{v,f} + \cdots,
\fe
 where $\cdots$ include higher level states, which are omitted in the level truncation approximation, as well as ghost-number 2 states that are annihilated by $b_0$ and therefore do not contribute to $\Psi_2$ (\ref{eq:Psi2Def}). The coefficients $A_{2,v,f}$ can be calculated using the cubic vertex conservation laws via
 \ie
\sum_{f'} A_{2,v,f'} (f,f') = \vv{v^c,f | \widetilde{\Psi}_2} = \{|v^c,f\rangle, \Psi_1, \Psi_1 \}.
 \fe
For $\Psi_1 = c \cosh(X^0-u)$, it follows from the free boson OPE that the only non-vanishing coefficients $A_{2,v,f}$ appear with the function $f$ of the form $f(X^0) \in \{ 1, ~ \cosh(2(X^0 - u)),~ \sinh(2(X^0 - u))\}$. Note that $\widetilde{\Psi}_2$ is also parity-even, and so $A_{2,v,f}$ is nonzero for even $n_M$ (the total number of $\alpha$ oscillators) in the cases $f=1, ~ \cosh(2(X^0 - u))$, and for odd $n_M$ in the case $f= \sinh(2(X^0 - u))$.

To calculate $\Omega^{(4)}$, we also need 
 \ie
 \Psi_3 = -\frac{b_0}{L_0} \widetilde{\Psi}_3, ~~~~~~~ \widetilde{\Psi}_3 \equiv  \Psi_1*\Psi_2 + \Psi_2 * \Psi_1 ,
 \fe
 where $\widetilde{\Psi}_3$ takes the form
  \ie\label{psithreet}
 \ket{\widetilde{\Psi}_3} =  \sum_{v,f, |v| \leq {\bf L}} A_{3,v,f} c_0 \ket{v,f} + \cdots.
 \fe
The non-vanishing $A_{3,v,f}$ appears with $f(X^0) \in \{\cosh(X^0 - u), ~ \sinh(X^0 - u), ~ \cosh(3(X^0 - u)), ~ \sinh(3(X^0 - u))\}$. Note that importantly, as mentioned in section \ref{sec:solution}, the exact marginality of the corresponding BCFT deformation implies that $\tilde{\Psi}_3$ cannot contain weight zero states, as the latter would otherwise obstruct the perturbative solution to the string field equation. In particular, a term proportional to $c_0 \ket{\Psi_1}=\ket{1}_\text{Liouville} \otimes \ket{\cosh(X^0-u)}_{X^0} \otimes c_0 \ket{\downarrow}_{bc}$ cannot appear in (\ref{psithreet}).\footnote{At finite truncation level ${\bf L}$ however, a term proportional to $c_0 \ket{\Psi_1}$ may appear in (\ref{psithreet}) with a small coefficient, but will nonetheless be omitted ``by hand'' in our truncation scheme.}

Having calculated $\Psi_2$ and $\Psi_3$, we can express $\Omega^{(4)}$ via (\ref{OSFTsymp}),(\ref{defD}) as
\ie
\label{Omega4}
\Omega^{(4)} = \frac{\lambda^3 \delta u \delta  \lambda}{2g_o^2} \Big\langle D(i) \Big[&f_1 \circ \Psi_1 ~ f_2 \circ \partial_u \Psi_3 - 3 f_1 \circ \partial_u \Psi_1 ~ f_2 \circ  \Psi_3 
\\
& + 3 f_1 \circ \Psi_3 ~ f_2 \circ \partial_u \Psi_1 -  f_1 \circ \partial_u \Psi_3 ~ f_2 \circ  \Psi_1
\\
& + 2 f_1 \circ \Psi_2 ~ f_2 \circ \partial_u \Psi_2 - 2 f_1 \circ \partial_u \Psi_2 ~ f_2 \circ  \Psi_2\Big] \Big\rangle^{\rm UHP} .
\fe
As $\Psi_2$ and $\Psi_3$ are not primaries, their conformal transformations with respect to the maps $f_1, f_2$ are not simple. It is convenient to evaluate these correlators using Ward identities similar to those of the cubic vertex conservation laws, which are described in detail in appendix \ref{app:sympCons}. In the end, a correlator of the form $\vv{D(i) ~ f_1 \circ A(w_1=0) ~ f_2 \circ B(w_2=0)}^{\text{UHP}}$ can be reduced to a sum of correlators of the same form but with $A$ and $B$ primaries, which are then easily evaluated (cf. \cite{Polchinski:1998rq}).

In applying the conservation laws (\ref{alphaLaw} - \ref{L25Law}) to (\ref{Omega4}), we encounter several simplifications. Firstly, using (\ref{cLaw}), (\ref{bLaw}), we can reduce every correlator involved to those with just three $c$-ghost modes:
(i) one each at $z= i,\, 1,\, -1$, (ii) two modes at $z=1$ and one at $z=-1$, (iii) one at $z=1$ and two modes at $z=-1$.
The latter two cases evaluate to zero, because $c_{-n} c_1$ acting on the insertion at $z=1$ can be turned to $c_n c_1$ acting on the insertion at $z=-1$. Only the case (i) can contribute to the symplectic form. Secondly, $L^{(25)}$ modes act trivially on $D(i)$, and so the $L^{(25)}$ modes at $z=\pm1$ produce the usual Gram matrix elements.
Finally,  $\Psi_1$ is itself a conformal primary, and so any creation mode moved onto $\Psi_1$ will annihilate it. Consequently, a correlator involving $\Psi_1$ can contribute only when the $\Psi_1$ is paired with terms in $\Psi_3$ that come with a single $\alpha_{-m}$ mode, which is then moved to act on the insertion at $z=i$ rather than on $\Psi_1$ at $z= \pm 1$. In other words, the only states in $\ket{\Psi_3}$ that contribute to the symplectic form are of the form $ \ket{1}_\text{Liouville} \otimes \alpha_{-m} \ket{\sinh(X^0-u)}_{X^0} \otimes \ket{\downarrow}_{bc}$. After applying these relations, the level-truncated $\Omega^{(4)}$ reduces to a linear combination of the following correlators:
\ie {}
& C_{0,cc}(k)\equiv \frac{1}{V_{X^0}} \vv{c(\pm i ) ~f_1 \circ c \cosh(kX^0)~ f_2 \circ c \cosh(kX^0) }^{\text{doubling}}= \frac{1}{2} ,\\
& C_{0,ss}(k) \equiv \frac{1}{V_{X^0}}\vv{c(\pm i ) ~f_1 \circ c \sinh(kX^0)~ f_2 \circ c \sinh(kX^0) }^{\text{doubling}} = -\frac{1}{2} ,\\
&C_{1,cs}(k)\equiv \frac{1}{V_{X^0}}\vv{c \partial X^0_L(\pm i ) ~f_1 \circ c \cosh(kX^0)~ f_2 \circ c \sinh(kX^0) }^{\text{doubling}} =  \frac{k}{2} ,\\
&C_{1,sc}(k) \equiv \frac{1}{V_{X^0}}\vv{c \partial X^0_L(\pm i ) ~f_1 \circ c \sinh(kX^0)~ f_2 \circ c \cosh(kX^0) }^{\text{doubling}}  = -\frac{k}{2} ,
 \fe
where $k=1,2$.

\begin{figure}[h!]
\centering
\includegraphics[width=0.9\textwidth]{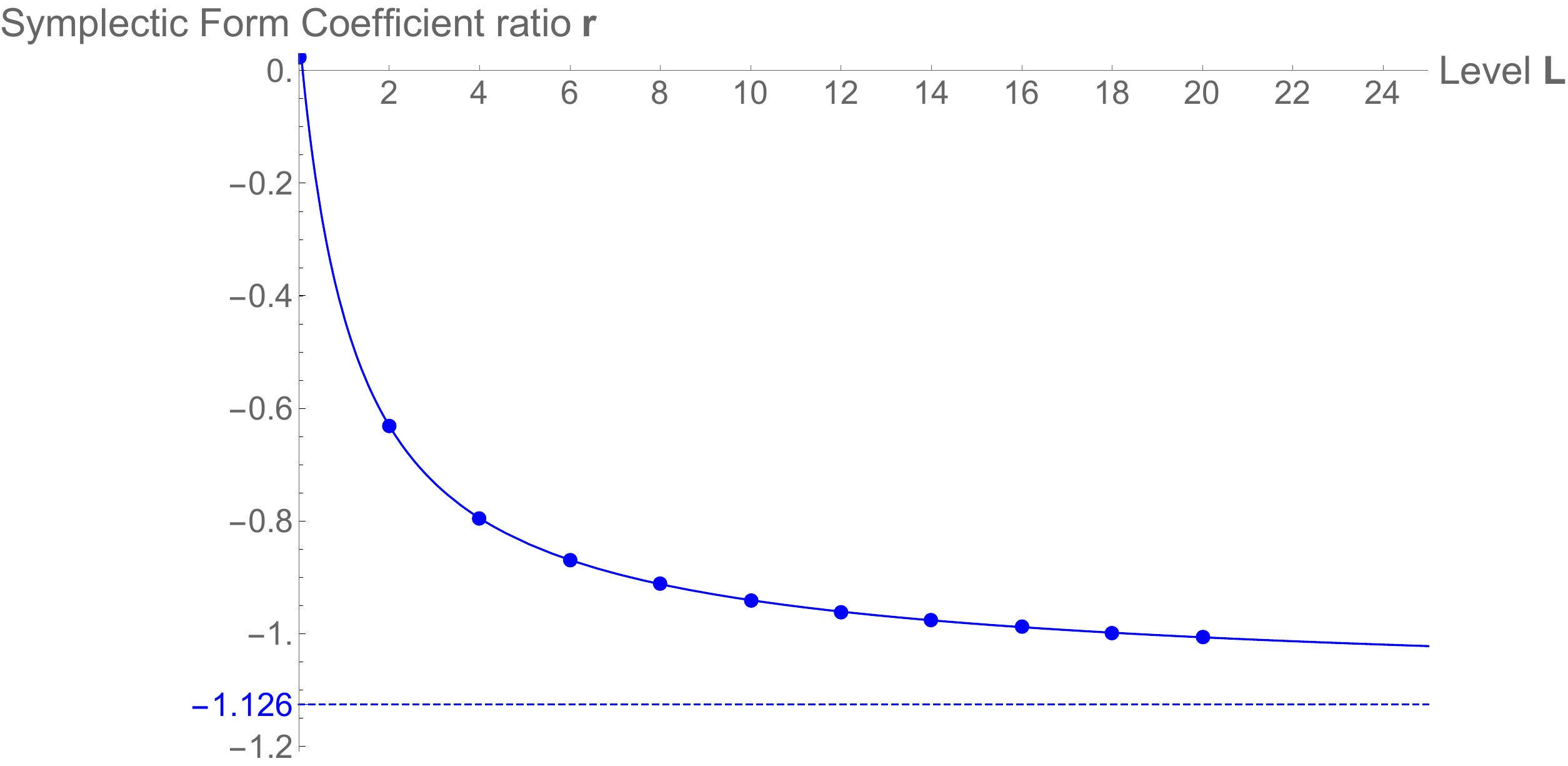}
\caption{ Results for $r$ from the level truncated symplectic form (blue dots), along with best fit of the form $c + \frac{a}{({\bf L}-d)^b}+ \frac{a_1}{({\bf L}-d)^{b+1}}$ (blue solid line), where ${\bf L}$ is the truncation level. The ${\bf L}\to \infty$ limit value is $r(=c)=-1.126$ (blue dotted line). }
\label{fig:sympExtrapolation}
\end{figure}

As a sample of the output of the level truncation computation, the lowest level results for the symplectic form are\footnote{The level truncation results are identical between ${\bf L} = 2n$ and ${\bf L} = 2n+1$ with $n=0,1,2, \hdots$, and so we only present the even ${\bf L}$ cases.}
\ie
\Omega^{(4)}|_{{\bf L}=0} &\approx  \frac{\lambda^3 \delta u  \delta  \lambda}{2g_o^2} 
\Big[  0.02312  (2C_{1,sc}(2)) - 0.02312  (2C_{1,cs}(2)) \Big] \approx  \delta u \,\delta \left(\frac{ (-0.09249)}{8g_o^2} \lambda^4 \right), 
\\
\Omega^{(4)}|_{{\bf L}=2} &\approx \frac{\lambda^3 \delta u  \delta  \lambda}{2g_o^2}  
\Big[ 0.02986  (2C_{1,sc}(2))- 0.02986  (2C_{1,cs}(2)) + 0.01233 (2C_{0,cc}(2)) \\
&\qquad\qquad - 0.01233 (2C_{0,ss}(2)) - 0.66680 (2C_{0,cc}(1))+ 2.0004 (2C_{0,ss}(1)) \Big] \\
& \approx \delta u \,\delta \left(\frac{ (0.63077)}{8g_o^2} \lambda^4\right).
\fe
We numerically evaluated $\Omega^{(4)}$ using level truncation up to ${\bf L}=20$. The results, shown in Figure \ref{fig:sympExtrapolation}, are then extrapolated ${\bf L } \to \infty$ by fitting a function of ${\bf L}$ of the form $c + \frac{a}{({\bf L}-d)^b}++ \frac{a_1}{({\bf L}-d)^{b+1}}$ to give an estimated value
\be
\label{eq:rFromSymp}
r \equiv {E_4\over E_2} \approx -1.126~~~~ ({\rm from~symplectic~form.})
\ee

\subsection{Comparison to Sen's proposal}\label{sec:comparisonToSen}

In order to compare the result (\ref{eq:rFromSymp}) with Sen's expression for the energy of the rolling tachyon based on the BCFT (\ref{SenEnergy}),
\be
E(\lambda(\tilde{\lambda})) = \frac{1}{2\pi^2 g_o^2} \left(1 - \pi^2 \tilde{\lambda}^2 +  \frac{\pi^4}{3} \tilde{\lambda}^4 + \hdots \right),
\ee
we still need to know the precise relation between the BCFT parameter $\tilde{\lambda}$ and the OSFT parameter $\lambda$. This relation is determined through (\ref{ellwoodDilaton}), where the Ellwood invariant of the OSFT solution can be evaluated perturbatively in $\lambda$ via (\ref{pertSol}). 

Previously, for the analogous exactly marginal BCFT deformation of a compact free boson at the self-dual radius, the Ellwood invariant  (\ref{ellwoodDilaton}) for the perturbative OSFT solution generated by $  \ket{\Psi_1}=\ket{1}_\text{Liouville} \otimes \ket{\cos X}_{X} \otimes c_0 \ket{\downarrow}_{bc}$ was calculated using level truncation in \cite{Kudrna:2019xnw} (section 6). Via Wick rotation, an essentially identical calculation applies to the rolling tachyon of present interest. A small technical issue is that \cite{Kudrna:2019xnw} had adopted a slightly different level truncation scheme, where the cutoff is imposed on the total weight
\be
L_v^{\rm K} = |K|+|M|+|N|+|P|+h_f = L_v + h_f
\label{eq:levelDef}
\ee 
of the state $|v,f\rangle$ (defined below (\ref{eq:Psi2Def})), rather than the oscillator level $L_v$. Here $h_f$ is the weight of the boundary-normal-ordered operator $f(X)$. This is sensible in the compact boson case, where the non-negative-weight operators $e^{ikX}$ give rise to a complete basis. Such a basis is not suitable, a priori, in the case of the noncompact, timelike free boson of consideration here. Furthermore, a cutoff that involves the weight of $f(X^0)$ appearing in the string field would be incompatible with locality in time. For these conceptual reasons, we have re-evaluated the Ellwood invariant appearing in (\ref{ellwoodDilaton}) by truncating with respect to the oscillator level $L_v$, and verified that at sufficiently high truncation levels the deviation from the numerical results of  \cite{Kudrna:2019xnw} is negligible.

\begin{figure}[h!]
\centering
\includegraphics[width=0.9\textwidth]{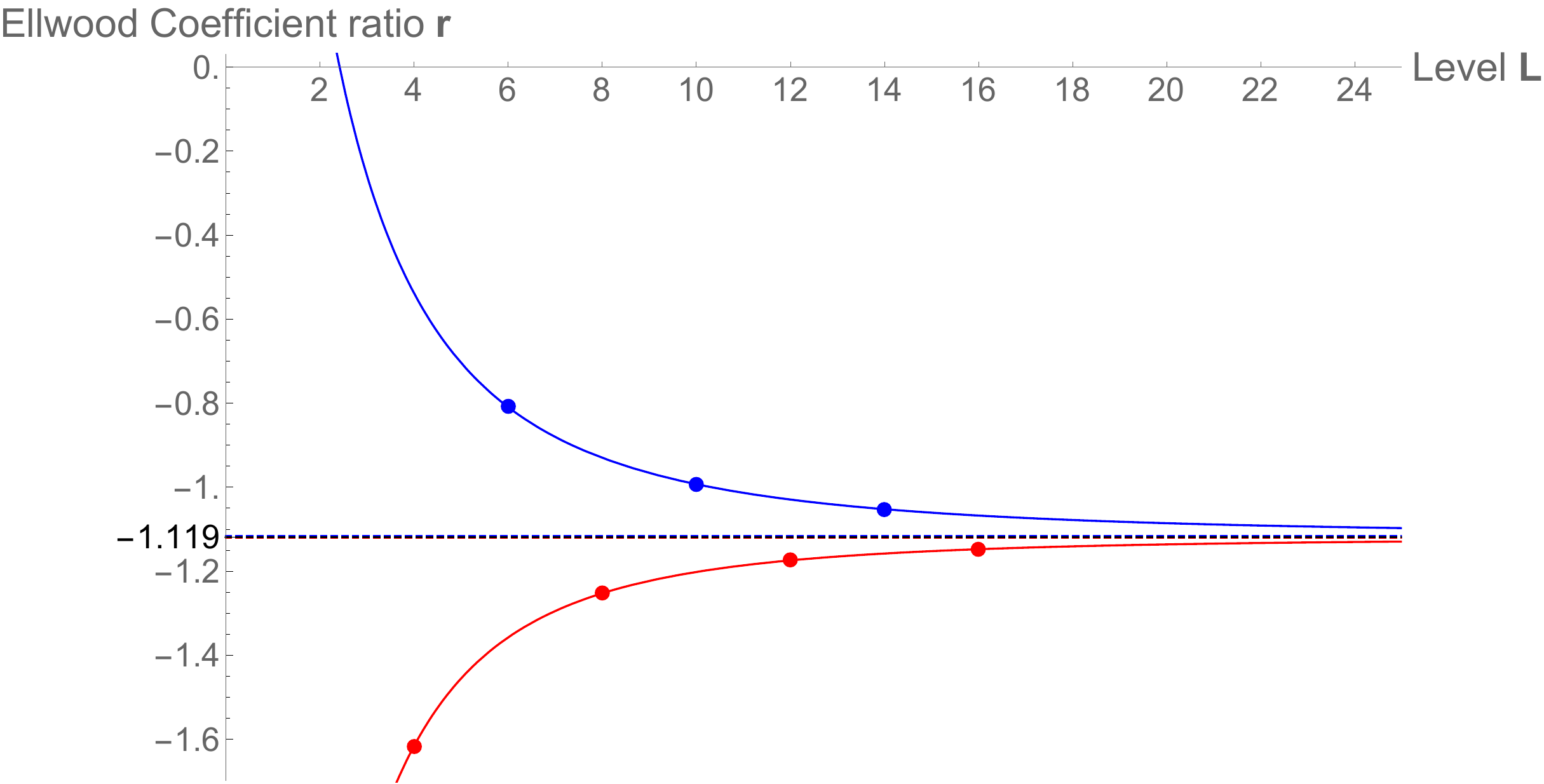}
\caption{ Results for $r$  from the level truncated Ellwood invariant (blue and red dots), along with best fits of the form $c + \frac{a}{({\bf L}-d)^b}$, where ${\bf L}$ is the truncation level. The ${\bf L}\to \infty$ limit values are $r(=c)=-1.117$ (blue dotted line) and  $r(=c)=-1.121$ (red dotted line). The average of the two fits as ${\bf L} \to \infty$ is  $r(=c)=-1.119$ (black dotted line)}
\label{fig:ellwoodExtrapolation}
\end{figure}

Writing
\be
{4\pi i\over V_{X^0}}W(\Psi(\tilde\lambda,u),V_D)=\text{cos}(2\pi\tilde\lambda) =  1 + D_1^{(2)} \lambda^2 + D_1^{(4)} \lambda^4 + \cdots,
\ee
where the exact value of $D_1^{(2)}$ is $-2\pi^2$, we have calculated $D_1^{(2)}$ and $D_1^{(4)}$ up to truncation level ${\bf L}=16$. Numerical extrapolation of $r=D_1^{(4)}/D_1^{(2)}$ to ${\bf L} = \infty$ using a fitting of the form $c + \frac{a}{({\bf L}-d)^b}$ (Figure \ref{fig:ellwoodExtrapolation}) gives the result
\be
r \equiv {D_1^{(4)}\over D_1^{(2)}} \approx -1.119 ~~~~  \text{ (from BCFT/Ellwood invariant.)}
\label{eq:rFromEllwood}
\ee

The result (\ref{eq:rFromSymp}), derived from the symplectic form of the OSFT, and (\ref{eq:rFromEllwood}) motivated by consideration of closed strings \cite{Sen:2004zm,Sen:2004cq,Sen:2004yv} and calculated through the BCFT, are in good agreement. The small discrepancy between (\ref{eq:rFromSymp}) and (\ref{eq:rFromEllwood}) can be attributed to the numerical extrapolation to ${\bf L} = \infty$ as well as errors in the level truncation approximation itself. Nonetheless, we believe that these level truncation results constitute a sufficiently nontrivial piece of evidence for the agreement between the two conceptually different prescriptions of the energy of the rolling tachyon, and furthermore provides a consistency test of all-order analytic arguments of section \ref{sec:energyanalytic}.

\section{An analytic evaluation of the symplectic form}
\label{sec:energyanalytic}

In this section, we will present an analytic argument that the energy of the rolling tachyon as determined from the symplectic form $\Omega$ (\ref{OSFTsymp}) agrees with the Ellwood invariant  (\ref{ellwoodDilaton}) (modulo the volume factor) of the string field solution $\Psi({\tilde\lambda},u)$ derived from the deformed BCFT.

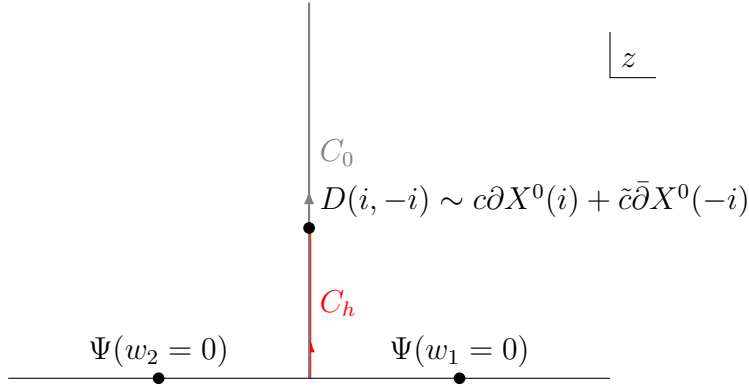
\begin{figure}[h]
\centering
\begin{tikzpicture}[scale=2]
\begin{scope}
\clip (1.9,1.9) rectangle (2.3,2.3);
\draw (2,2) node [anchor=south west] {$z$} rectangle (2.5,2.5) ;
\end{scope}
\begin{scope}
[decoration={
    markings,
    mark=at position .5  with {\arrow[gray]{latex}}}
    ]
\path[draw=gray,thick,postaction={decorate}] (0,0) -- (0,2.5);
\node at (0,1.5) [anchor=west,color=gray] {$C_0$};
\end{scope}
\begin{scope}
[decoration={
    markings,
    mark=at position .25  with {\arrow[red,very thin]{Latex[harpoon,swap]}}}
    ]
\path[draw=red,thin,postaction={decorate}] (.3pt,0) -- (.3pt,1);
\node at (0,0.5) [anchor=west,color=red] {$C_h$};
\end{scope}
\path[draw] (-2,0) -- (2,0);
\draw[fill=black] (0,1) circle [radius=1pt] node [anchor=south west] {$D(i,-i)\sim c\partial X^0(i) + \tilde{c} \bar{\partial} X^0(-i)$};
\draw[fill=black] (1,0) circle [radius=1pt] node [anchor=south] {$\Psi(w_1=0)$};
\draw[fill=black] (-1,0) circle [radius=1pt] node [anchor=south] {$\Psi(w_2=0)$};
\end{tikzpicture}
\caption{The correlator on the UHP computing $\Omega$ in (\ref{omegaUHP}).}
\end{figure}

To begin with, we observe that the time derivative of the string field solution inserted at a point $y$ on the boundary (real axis) of the UHP can be expressed as
\ie{}
{\D\over\D u}\Psi({\tilde\lambda},u)(y)&=-{\D\over\D X^0}\Psi({\tilde\lambda},u)(y)
\\
&=-\int_{C_y}\left({dz\over2\pi i}\partial X^0(z)-{d{\bar z}\over2\pi i}{\bar\partial}X^0({\bar z})\right)\cdot\Psi({\tilde\lambda},u)(y)
\\
&\equiv-A_{C_y}\cdot\Psi({\tilde\lambda},u)(y).
\fe
Here $C_y$ is a counter-clockwise semi-circle contour enclosing $y$, that can be freely deformed on the UHP as long as it does not pass through other insertions of {\it zero modes} of $X^0$. In particular, we may move this contour across $D(i)$ (\ref{defD}) appearing in the definition of $\Omega$ (\ref{OSFTsymp}), and hence write the symplectic form as
\ie\label{omegaUHP}
\Omega
&={\D u\D\tilde\lambda\over2g_o^2}\bigg\langle D(i) ~\bigg({\D\over\D\tilde\lambda}f_1\circ\Psi({\tilde\lambda},u)(w_1=0) ~{\D\over\D u}f_2\circ\Psi({\tilde\lambda},u)(w_2=0)
\\
&~~~~-{\D\over\D u}f_1\circ\Psi({\tilde\lambda},u)(w_1=0) ~{\D\over\D \tilde\lambda}f_2\circ\Psi({\tilde\lambda},u)(w_2=0)\bigg)\bigg\rangle^{\rm UHP}
\\
&=-{\D u\D\tilde\lambda\over2g_o^2}{\D\over\D\tilde\lambda}\bigg\langle A_{C_0} D(i) ~f_1\circ\Psi({\tilde\lambda},u)(w_1=0) ~f_2\circ\Psi({\tilde\lambda},u)(w_2=0)\bigg\rangle^{\rm UHP},
\fe
where $C_0$ is a contour that extends along the entire positive imaginary axis.

Via the relation $\Omega = \delta u\delta\tilde\lambda { \delta E\over\delta\tilde\lambda}$, we can re-express this in terms of the variation of the energy $E(\tilde\lambda)$. Next, by folding the contour $C_0$ in half, and employing the doubling trick, we have
\ie\label{UHPtoDoubl}
{\D E\over\D\tilde\lambda} 
&=-{1\over g_o^2}{\D\over\D\tilde\lambda}\bigg\langle A_{C_h} D(i) ~f_1\circ\Psi({\tilde\lambda},u)(w_1=0) ~f_2\circ\Psi({\tilde\lambda},u)(w_2=0)\bigg\rangle^{\rm UHP}
\\
&=-{1\over g_o^2 V_{X^0}}{\D\over\D\tilde\lambda}\bigg\langle {\tilde A}_{C_f} (c\partial X_L^0(i)+c{\partial}X_L^0(-i)) ~f_1\circ\Psi({\tilde\lambda},u)(w_1=0) f_2\circ\Psi({\tilde\lambda},u)(w_2=0)\bigg\rangle^{\rm doubling},
\fe
where $C_h$ is the contour extended from the origin to $i$ along the imaginary axis, and $C_f$ is the contour extended from from $-i$ to $i$ related by doubling trick, with 
\ie
{\tilde A}_{C_f} \equiv \int_{C_f}{dz\over2\pi i}\partial X^0_L(z).
\fe
The integration along $C_f$ can be performed explicitly, with the result expressed in terms of $X_L$ insertions at the end points $z=\pm i$, giving
\ie\label{precorrl}
{\D E\over\D\tilde\lambda}&={i\over 2\pi g_o^2V_{X^0}}{\D\over\D\tilde\lambda}\bigg\langle \left(cX_L^0\partial X_L^0(i)-{c}X_L^0{\partial}X_L^0(-i)+X_L^0(i){c}{\partial}X_L^0(-i)-X_L^0(-i)c\partial X_L^0(i)  \right) ~
\\
&~~~~~~~~~~~~~~~~~~~~~~~~~~~~~~~~~~~~f_1\circ\Psi({\tilde\lambda},u)(w_1=0) ~f_2\circ\Psi({\tilde\lambda},u)(w_2=0)\bigg\rangle^{\rm doubling}.
\fe

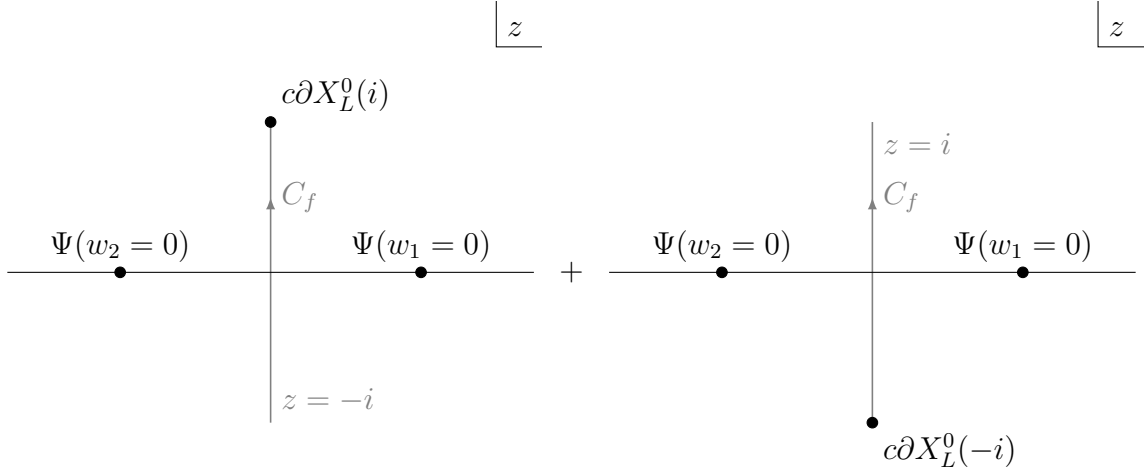
\begin{figure}[h]
\centering
\begin{tikzpicture}[scale=2]
\begin{scope}[xshift=-2cm]
\begin{scope}
\clip (1.4,1.4) rectangle (1.8,1.8);
\draw (1.5,1.5) node [anchor=south west] {$z$} rectangle (1.9,1.9) ;
\end{scope}
\begin{scope}
[decoration={
    markings,
    mark=at position .75  with {\arrow[gray]{latex}}}
    ]
\path[draw=gray,semithick,postaction={decorate}] (0,-1) -- (0,1);
\node at (0,0.5) [anchor=west,color=gray] {$C_f$};
\node at (0,-1) [anchor=south west,color=gray] {$z=-i$};
\end{scope}
\path[draw] (-1.75,0) -- (1.75,0);
\draw[fill=black] (0,1) circle [radius=1pt] node [anchor=south west] {$c\partial X_L^0(i)$};
\draw[fill=black] (1,0) circle [radius=1pt] node [anchor=south] {$\Psi(w_1=0)$};
\draw[fill=black] (-1,0) circle [radius=1pt] node [anchor=south] {$\Psi(w_2=0)$};
\end{scope}

\begin{scope}[xshift=2cm]
\begin{scope}
\clip (1.4,1.4) rectangle (1.8,1.8);
\draw (1.5,1.5) node [anchor=south west] {$z$} rectangle (1.9,1.9) ;
\end{scope}
\begin{scope}
[decoration={
    markings,
    mark=at position .75  with {\arrow[gray]{latex}}}
    ]
\path[draw=gray,semithick,postaction={decorate}] (0,-1) -- (0,1);
\node at (0,0.5) [anchor=west,color=gray] {$C_f$};
\node at (0,1) [anchor=north west,color=gray] {$z=i$};
\end{scope}
\path[draw] (-1.75,0) -- (1.75,0);
\draw[fill=black] (0,-1) circle [radius=1pt] node [anchor=north west] {$c\partial X_L^0(-i)$};
\draw[fill=black] (1,0) circle [radius=1pt] node [anchor=south] {$\Psi(w_1=0)$};
\draw[fill=black] (-1,0) circle [radius=1pt] node [anchor=south] {$\Psi(w_2=0)$};
\end{scope}

\node at (0,0) {$+$};
\end{tikzpicture}
\caption{Correlator (\ref{UHPtoDoubl}) after the doubling trick.}
\end{figure}

In fact, the part of the correlator on the RHS of (\ref{precorrl}) with insertion of $cX_L^0\partial X_L^0$ at $z=i$ vanishes. To see this, we first replace the insertion of the two string fields $\Psi$ at $z=\pm 1$ with their star product $\Psi*\Psi$ inserted at the origin $z=0$, transformed with the conformal map (\ref{zfwmap}). Next, we use the equation of motion obeyed by $\Psi$, $\Psi*\Psi=-Q_B\Psi$, and deform the BRST contour to enclose the operator inserted at $z=i$. This results in the identity
\ie\label{confTransfJ}
& \bigg\langle cX_L^0\partial X_L^0(i)~f_1\circ\Psi({\tilde\lambda},u)(w_1=0)~f_2\circ\Psi({\tilde\lambda},u)(w_2=0)\bigg\rangle^{\rm doubling}
\\
&=  \bigg\langle \left(cX_L^0\partial X_L^0\right)(i)~f\circ\left(\Psi({\tilde\lambda},u)*\Psi({\tilde\lambda},u)\right)(w=0)\bigg\rangle^{\rm doubling}
\\
&=  \bigg\langle Q_B\cdot\left(cX_L^0\partial X_L^0\right)(i)~f\circ\Psi({\tilde\lambda},u)(w=0)\bigg\rangle^{\rm doubling}
\\
&=  \bigg\langle \left(-{1\over4}c\partial^2c\right)(i)~f\circ\Psi({\tilde\lambda},u)(w=0)\bigg\rangle^{\rm UHP}.
\fe
The correlator appearing in the last line is also known as the Ellwood invariant $W\left(\Psi({\tilde\lambda},u),-{1\over4}c\partial^2c\right)$, as the bulk operator $c\partial^2c$ is $Q_B$-closed. In this case, the bulk insertion is pure ghost whereas the string field solution $\Psi$ amounts to an exactly marginal deformation of the matter $X^0$ CFT, by (\ref{ellwoodConj}) the result vanishes identically. A similar argument applies to the part of  the correlator on the RHS of (\ref{precorrl}) with $cX_L^0\partial X_L^0$ inserted at $z=-i$. We are thus left with
\ie\label{Ederiv}
{\D E\over\D\tilde\lambda}&={i\over 2\pi g_o^2V_{X^0}}{\D\over\D\tilde\lambda}\bigg\langle \left(X_L^0(i){c}{\partial}X_L^0(-i)-X_L^0(-i)c\partial X_L^0(i)  \right) ~
\\
&~~~~~~~~~~~~~~~~~~~~~~~~~~~~~~~~~~~~f_1\circ\Psi({\tilde\lambda},u)(w_1=0) ~f_2\circ\Psi({\tilde\lambda},u)(w_2=0)\bigg\rangle^{\rm doubling}.
\fe

Let us compare (\ref{Ederiv}) with the $\tilde\lambda$-variation of the Ellwood invariant (\ref{ellwoodDilaton}) associated with the bulk zero momentum dilaton operator $V_D$. Using 
\ie
V_D=-{1\over2}Q_B\cdot\left(R+{1\over4}\partial c-{1\over4}{\bar\partial}{\tilde c} \right),~~~~R\equiv X^0\left(c\partial X^0-{\tilde c}{\bar\partial}X^0 \right),
\fe
we can write
\ie\label{Wderiv}
&{\D\over\D\tilde\lambda} W(\Psi(\tilde\lambda,u),V_D)
\\
&=-{1\over2}{\D\over\D\tilde\lambda}\left(W(\Psi(\tilde\lambda,u),Q_B\cdot R)+{1\over4}W(\Psi(\tilde\lambda,u),Q_B\cdot{\partial c})-{1\over4}W(\Psi(\tilde\lambda,u),Q_B\cdot{\bar\partial}{\tilde c}) \right).
\fe
Via the doubling trick, we can replace $R$ inserted at $z=i$ on the UHP with 
\ie
R^{\rm doubling} = cX_L^0\partial X_L^0(i)-cX_L^0\partial X_L^0(-i)-X_L^0(i)c\partial X_L^0(-i)+X_L^0(-i)c\partial X_L^0(i).
\fe
Further using $cX_L^0\partial X_L^0={1\over2}Q_B\left(X_L^0\right)^2-{1\over4}\partial c$, the contribution from the first two terms in $R^{\rm doubling}$ to (\ref{Wderiv}) cancels against that of the last two terms in (\ref{Wderiv}). Now deforming the BRST contour and using $-Q_B\Psi=\Psi*\Psi$, we arrive at
\ie\label{Wderiv2}
{i\over\pi g_o^2 V_{X^0}}{\D\over\D\tilde\lambda} W(\Psi(\tilde\lambda,u),V_D)
&={i\over2\pi g_o^2V_{X^0}}{\D\over\D\tilde\lambda}\bigg\langle \left(X_L^0(i){c}{\partial}X_L^0(-i)-X_L^0(-i)c\partial X_L^0(i)  \right) ~
\\
&~~~~~~~~~~~~~~~~f_1\circ\Psi({\tilde\lambda},u)(w_1=0) ~f_2\circ\Psi({\tilde\lambda},u)(w_2=0)\bigg\rangle^{\rm doubling}.
\fe
The agreement of this expression with ${\D E\over\D\tilde\lambda}$ in (\ref{Ederiv}) hence establishes the relation between the energy $E(\lambda(\tilde\lambda))$ and the Ellwood invariant $W(\Psi(\tilde\lambda,u),V_D)$, as claimed.

\begin{figure}[h]
\centering
\begin{tikzpicture}[scale=1.5]
\begin{scope}[xshift=-2.25cm,yshift=2cm]
\begin{scope}
\clip (1.4,1.4) rectangle (1.8,1.8);
\draw (1.5,1.5) node [anchor=south west] {$z$} rectangle (1.9,1.9) ;
\end{scope}
\begin{scope}
[decoration={
    markings,
    mark=at position 0  with {\arrow[blue]{latex[flex']}}}
    ]
\path[draw=blue,semithick,postaction={decorate}] (0,1) circle [radius=20pt];
\node at ($(0,1)+(45: 20pt)$) [anchor=west,color=blue] {$Q_B$};
\end{scope}
\path[draw] (-1.75,0) -- (1.75,0);
\path[draw=gray,dashed,thick] (0,1) -- (0,2);
\draw[fill=black] (0,1) circle [radius=1pt] node [anchor=north] {$R+\frac{\partial c -\bar{\partial} \tilde{c}}{4}$};
\draw[fill=black] (0,0) circle [radius=1pt] node [anchor=north] {$\Psi(w=0)$};
\end{scope}

\path[draw, arrows = -Computer Modern Rightarrow] (-0.25cm,2cm) -- (0.25cm,2cm); 

\begin{scope}[xshift=2.25cm,yshift=2cm]
\begin{scope}
\clip (1.4,1.4) rectangle (1.8,1.8);
\draw (1.5,1.5) node [anchor=south west] {$z$} rectangle (1.9,1.9) ;
\end{scope}
\begin{scope}
[decoration={
    markings,
    mark=at position .25  with {\arrowreversed[blue]{latex[flex']}}}
    ]
\path[draw=blue,semithick,postaction={decorate}] (0,0) circle [radius=20pt];
\node at (45: 20pt) [anchor=west,color=blue] {$Q_B$};
\end{scope}
\path[draw] (-1.75,0) -- (1.25,0);
\draw[fill=black] (0,1) circle [radius=1pt] node [anchor=south west] {$X_L^0$};
\draw[fill=black] (0,-1) circle [radius=1pt] node [anchor=west] {$c \partial X_L^0$};
\draw[fill=black] (0,0) circle [radius=1pt] node [anchor=north] {$\Psi(w=0)$};
\end{scope}

\node at (3.75cm,2cm) {$-$};

\begin{scope}[xshift=5.25cm,yshift=2cm]
\begin{scope}
\clip (1.4,1.4) rectangle (1.8,1.8);
\draw (1.5,1.5) node [anchor=south west] {$z$} rectangle (1.9,1.9) ;
\end{scope}
\begin{scope}
[decoration={
    markings,
    mark=at position .25  with {\arrowreversed[blue]{latex[flex']}}}
    ]
\path[draw=blue,semithick,postaction={decorate}] (0,0) circle [radius=20pt];
\node at (45: 20pt) [anchor=west,color=blue] {$Q_B$};
\end{scope}
\path[draw] (-1.25,0) -- (1.75,0);
\draw[fill=black] (0,1) circle [radius=1pt] node [anchor=south west] {$c \partial X_L^0$};
\draw[fill=black] (0,-1) circle [radius=1pt] node [anchor=west] {$X_L^0$};
\draw[fill=black] (0,0) circle [radius=1pt] node [anchor=north] {$\Psi(w=0)$};
\end{scope}

\path[draw, arrows = -Computer Modern Rightarrow] (-3.25cm,-1.5cm) -- (-2.75cm,-1.5cm); 

\begin{scope}[xshift=-.75cm,yshift=-1.5cm]
\begin{scope}
\clip (1.4,1.4) rectangle (1.8,1.8);
\draw (1.5,1.5) node [anchor=south west] {$z$} rectangle (1.9,1.9) ;
\end{scope}
\path[draw] (-1.75,0) -- (1.25,0);
\draw[fill=black] (0,1) circle [radius=1pt] node [anchor=south west] {$X_L^0$};
\draw[fill=black] (0,-1) circle [radius=1pt] node [anchor=west] {$c \partial X_L^0$};
\draw[fill=black] (1,0) circle [radius=1pt] node [anchor=north east] {$\Psi(w_1=0)$};
\draw[fill=black] (-1,0) circle [radius=1pt] node [anchor=north east] {$\Psi(w_2=0)$};
\end{scope}

\node at (0.75cm,-1.5cm) {$-$};

\begin{scope}[xshift=2.25cm,yshift=-1.5cm]
\begin{scope}
\clip (1.4,1.4) rectangle (1.8,1.8);
\draw (1.5,1.5) node [anchor=south west] {$z$} rectangle (1.9,1.9) ;
\end{scope}
\path[draw] (-1.25,0) -- (1.75,0);
\draw[fill=black] (0,1) circle [radius=1pt] node [anchor=south west] {$c \partial X_L^0$};
\draw[fill=black] (0,-1) circle [radius=1pt] node [anchor=west] {$X_L^0$};
\draw[fill=black] (1,0) circle [radius=1pt] node [anchor=north west] {$\Psi(w_1=0)$};
\draw[fill=black] (-1,0) circle [radius=1pt] node [anchor=north west] {$\Psi(w_2=0)$};
\end{scope}

\end{tikzpicture}
\caption{Manipulations in the evaluation of (\ref{Wderiv}).}
\end{figure}
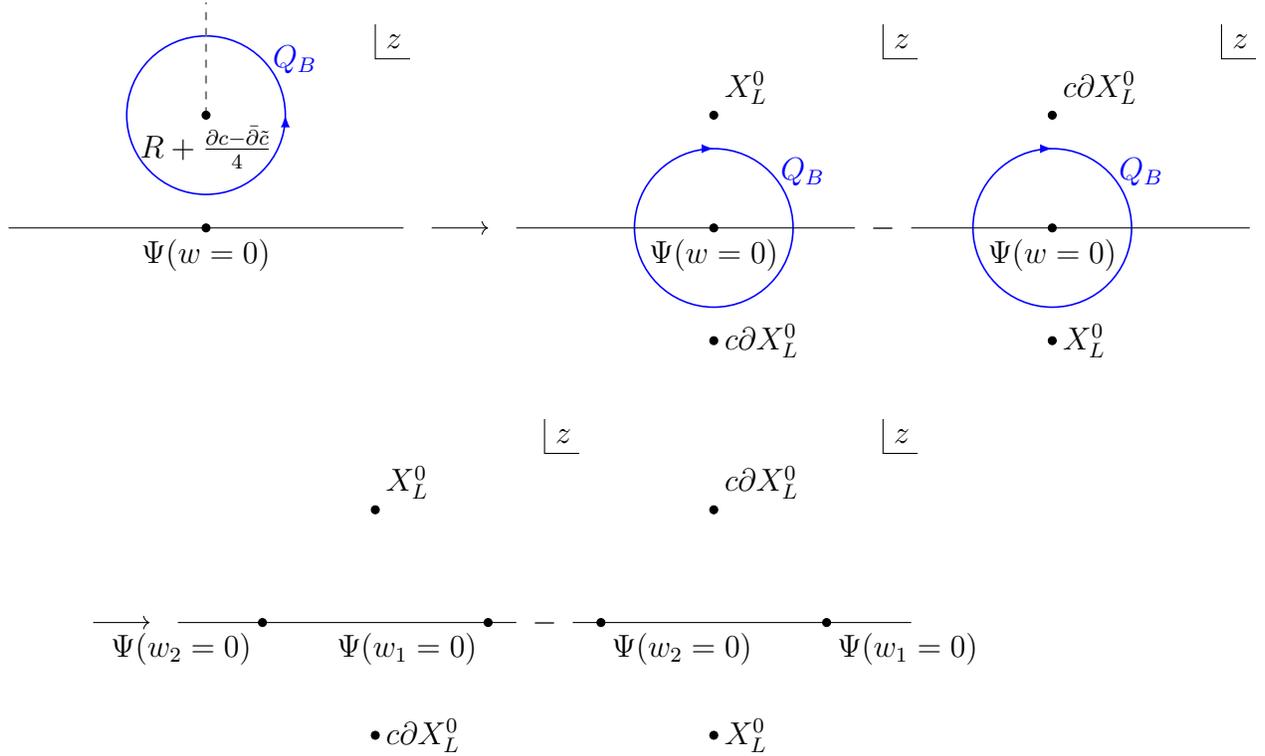

\section{Discussion}
\label{sec:discuss}

To summarize, we have shown that the energy of the rolling tachyon on the ZZ-brane in $c=1$ string theory, determined from the regularized symplectic form (\ref{OSFTsymp}) of the OSFT up to an overall constant shift, is equal to the Ellwood invariant of the zero momentum dilaton operator $V_D$, which in turn is the same as the coefficient of $V_D$ in the deformed rolling tachyon boundary state. The latter was previously given the interpretation of energy by Sen \cite{Sen:2002nu}, based on considerations of closed string backreaction  \cite{Sen:2002in,Sen:2004zm,Sen:2004yv}. In contrast, the derivation presented in this paper amounts to a direct construction of the OSFT Hamiltonian or equivalently its corresponding vector field on the phase space without appealing to coupling to closed strings.

Thus far, we have worked under the assumption that the relevant domain of the phase space of OSFT on the ZZ-brane is two-dimensional and consists entirely of the rolling tachyon solutions which can be constructed perturbatively. While this is suggested by the identification of the ZZ-brane with a fermion/eigenvalue of the dual matrix quantum mechanics \cite{McGreevy:2003kb, McGreevy:2003ep}, in addition to regions of the phase space that correspond to open string tachyon rolling to the ``wrong side'' (presumably described by similar OSFT solutions with negative or imaginary $\lambda$), there may be additional components of the phase space that captures multiple ZZ-branes or higher ZZ-branes defined by $(m,n)$ ZZ-boundary conditions \cite{Erler:2014eqa, Erler:2019fye}. While the higher ZZ-boundary conditions appear to play a role in D-instantons of $c=1$ string theory \cite{Balthazar:2019ypi}, the existence of the corresponding physical branes (that support open string tachyons) and their dual interpretation in the matrix quantum mechanics remain to be clarified. We hope the present work serves as a first step towards exploring the full phase space of the OSFT of ZZ-branes, and its quantization that may lead to the construction of a quantum non-perturbative OSFT (cf. \cite{Sen:2004nf}).

\section*{Acknowledgements}

We would like thank Carlo Maccaferri, Martin Schnabl, Ashoke Sen, and Charles Wang for discussions. We are especially grateful to Ted Erler for explaining to us subtleties in defining the symplectic form of OSFT \cite{Erler:2004hv}, and Matej Kudrna for sharing with us the details of level truncation data involved in \cite{Kudrna:2019xnw}. We thank Pedro Pascual Center of Science, Benasque, Spain where an early version of this work was presented. MC is supported by the Sam B. Treiman fellowship at the Princeton Center for Theoretical Science. BM and XY are supported by a Simons Investigator Award from the Simons Foundation, by the Simons Collaboration Grant on the Non-Perturbative Bootstrap, and by DOE grant DE-SC0007870. The numerical computations in this work are performed on the FAS Research Computing cluster at Harvard University.

\appendix

\section{OSFT conventions}
\label{sec:osftconvention}

In this Appendix, we summarize the OSFT conventions used in this paper, primarily based on those of \cite{Rastelli:2000iu,Kudrna:2019xnw}. In the classical OSFT, the open string field $\ket{\Psi}$ is a ghost number 1 state in the BCFT Hilbert space. The $\text{SL}(2,\mathbb{R})$-invariant vacuum $\ket{0}$ has ghost number zero, $b$ has ghost number $-1$ and $c$ has ghost number $+1$. Under the state-operator correspondence, $\ket{\Psi}$ maps to an operator $\Psi(w=0)$ inserted at the origin of the upper half unit disc (UHD). Propagated to the unit semicircle, the string has its midpoint at $w=i$, its left half at $\{w=e^{i\theta}|\; \theta \in [0,\pi] \}$ and right half at $\{w=e^{i\theta}|\; \theta \in [\pi,2\pi] \}$.

The star product between a pair of string fields, $\Psi_1*\Psi_2$, is defined by identifying the right half of $\Psi_1$ with the left half of $\Psi_2$. $*$ is associative, namely
\be
(\Psi_1 * \Psi_2) * \Psi_3 = \Psi_1 * (\Psi_2 * \Psi_3),
\ee
but not commutative nor anti-commutative.
The BRST charge $Q_B$ is a nilpotent, Grassmann-odd derivation on this algebra, that satisfies the Leibnitz rule
\be
Q_B (A * B) = (Q_B A) * B + (-1)^{A}A * (Q_B B),
\ee
where $A= \Psi$ or $A= \delta \Psi$ with $\delta$ defined in section \ref{sec:symplectic}. $(-1)^{A} = 1$ for Grassmann even $A$ and $-1$ for Grassmann odd $A$ (we adopt the convention that $\delta$ itself is Grassmann odd). The BRST charge $Q_B$ is given by
\ie
Q_B &= \int_C \frac{dz}{2\pi i} j_B - \frac{d\bz}{2 \pi i} \tilde{j}_B, \\
j_B &= cT^{\text{matter}}+ bc \partial c + \frac{3}{2} \partial^2 c ,
\fe
where $C$ is a counter-clockwise (CCW) semicircle contour enclosing the origin.

The convolution $\int$ is a linear functional on the space of string fields. It satisfies $\int Q_B \Psi = 0$ up to possible contributions from the boundaries of spacetime. It also satisfies the graded cyclicity property
\be
\int A * B = (-1)^{AB} \int B * A,
\ee
and by extension
\be
\int A_1 * A_2 * \cdots * A_n = (-1)^{(A_2+\hdots+A_n)A_1} \int A_2 * \cdots* A_n * A_1.
\ee
Such an $n$-linear functional on string fields is referred to as an $n$-vertex. The action of Witten's cubic OSFT (\ref{OSFTaction}) consists of 2- and 3-vertices, and the Ellwood invariant will be defined through a 1-vertex.

The $n$-vertex is calculated by a CFT correlator of $n$-string fields on the unit disc or the UHP, with specific choices of local charts containing each string field insertion. For instance, the 2-vertex is equivalent to the BPZ inner product
\ie
\int \Psi_1 * \Psi_2 = \vv{\Psi_1 | \Psi_2}.
\fe
The cubic vertex is defined by the correlator
\ie
\int \Psi_1 * \Psi_2 * \Psi_3 = \vv{\Psi_1,  \Psi_2, \Psi_3} = \vv{ g_1 \circ \Psi_1(w_1=0) ~ g_2 \circ \Psi_2(w_2=0) ~ g_3 \circ \Psi_3(w_3=0)}^{\text{unit disc}}
\fe
where
\ie
\begin{cases}
g_1(w_1) &= e^{\frac{2\pi i }{3 }} \left( \frac{1+iw_1}{1-iw_1} \right)^{\frac{2}{3}}\\[1pt]
g_2(w_2) &=\left( \frac{1+iw_2}{1-iw_2} \right)^{\frac{2}{3}}\\[1pt]
g_3(w_3) &= e^{-\frac{2\pi i }{3 }} \left( \frac{1+iw_3}{1-iw_3} \right)^{\frac{2}{3}}
\end{cases}
\fe
each maps the $w_i$-UHD to a wedge spanning angle $2\pi/3$ in the unit disc parameterized by $z'$ (with $|z'|\leq1$). The string field insertion at $w_i=0$ is mapped to one at the midpoint on the boundary arc of each wedge, and the midpoint of each string at $w_i = i$ is mapped to the center of the disc $z'=0$. The semicircle  $\{|w_i|=1,\, \Im\{w_i\} \geq 0\}$, on the other hand, is mapped to a pair of radii on the boundary of the wedge. In this formulation, the cyclicity of the cubic vertex is manifest.

\begin{figure}[h]
\centering
\subfloat[The cubic vertex on a disc]{
\begin{tikzpicture}
 \path[draw=red,thick] (-2,0.4pt) -- (0,.4pt);
 \path[draw=blue,thick] (-2,-0.4pt) -- (0,-.4pt);
 \path[draw=blue,thick] ($(0,0)+(150:0.4pt)$) -- ($(60:2)+(150:0.4pt)$);
 \path[draw=red,thick] ($(0,0)-(150:0.4pt)$) -- ($(60:2)-(150:0.4pt)$);
 \path[draw=blue,thick] ($(0,0)-(-150:0.4pt)$) -- ($(-60:2)-(-150:0.4pt)$);
 \path[draw=red,thick] ($(0,0)+(-150:0.4pt)$) -- ($(-60:2)+(-150:0.4pt)$);
 \draw (0,0) circle [radius=2];
 \draw[fill] (0,0) circle [radius=1pt];
 \draw[fill] (0:2) circle [radius=1pt];
 \draw[fill] (120:2) circle [radius=1pt];
 \draw[fill] (-120:2) circle [radius=1pt];
 \node (psi1) at (120:2) [label=above:$\Psi_1$]{};
 \node (psi2) at (0:2) [label=right:$\Psi_2$]{};
 \node (psi3) at (-120:2) [label=below:$\Psi_3$]{};
\end{tikzpicture}
}
\subfloat[The cubic vertex on the UHP]{
\begin{tikzpicture}[scale=2]
 \draw[color=blue,thick] plot[mark=none,smooth] file {plot1/plot1_1.txt};
 \draw[color=red,thick] plot[mark=none,smooth] file {plot1/plot1_2.txt};
 \draw[color=blue,thick] plot[mark=none,smooth] file {plot1/plot1_3.txt};
 \draw[color=red,thick] plot[mark=none,smooth] file {plot1/plot1_4.txt};
 \draw[color=red,thick] (0.2pt,1) -- (0.2pt,2.5);
 \draw[color=blue,thick] (-0.2pt,1) -- (-0.2pt,2.5); 
 \draw (-2,0) -- (2,0);
 \draw[fill] (0,0) circle [radius=0.5pt];
 \draw[fill] (0,1) circle [radius=0.5pt];
 \draw[fill] (-{sqrt(3)},0) circle [radius=0.5pt];
 \draw[fill] ({sqrt(3)},0) circle [radius=0.5pt];
 \node (psi1) at ({sqrt(3)},0) [label=below:$\Psi_1$]{};
 \node (psi2) at (0,0) [label=below:$\Psi_2$]{};
 \node (psi3) at (-{sqrt(3)},0) [label=below:$\Psi_3$]{};
\end{tikzpicture}
}\\
\subfloat[The UHD as a local chart containing a string field]{
\begin{tikzpicture}
 \begin{scope}
 \clip (-2.1,0) rectangle (0,2.1);
 \path[draw=blue,thick] (0,0) circle [radius=2];
 \end{scope}
 \begin{scope}
 \clip (0,0) rectangle (2.1,2.1);
 \path[draw=red,thick] (0,0) circle [radius=2];
 \end{scope}
 \draw (-2,0) -- (2,0);
 \draw[fill] (0,0) circle [radius=1pt];
 \node at (0,0) [label=above:$\Psi_1$]{};
\end{tikzpicture}
}
\end{figure}

The {\it identity string field} $\mathbb{I}$ is defined by the property 
\ie
\mathbb{I} * \Psi = \Psi * \mathbb{I} = \Psi
\fe
for all string fields $\Psi$. It follows that a 2-vertex involving $\mathbb{I}$ reduces to a 1-vertex,
\ie
\int  \Psi = \vv{ \mathbb{I} | \Psi}.
\fe
To express this 1-vertex as a CFT correlator on the unit disc or the UHP, we need a conformal transformation that maps the $w$-UHD (with $\Psi$ inserted at $w=0$) to the whole unit $z'$-disc in such a way that the semicircle  $\{|w|=1,\, \Im\{w\} \geq 0\}$ maps to a radius of the $z'$-disc, while the midpoint $w=i$ maps to the center of the disc $z'=0$. Note that the left and right halves of the $w$-semicircle are glued together on the $z'$-disc. The $z'$-disc can also be further mapped to the $z$-UHP, with the $w$-semicircle now folded into the semi-infinite line $\{z=iy\, | \, y \in [1,\infty) \}$. That is,
\ie
\int  \Psi = \vv{F_I \circ \Psi(w=0)}^{\text{unit disk}} = \vv{f_I \circ \Psi(w=0)}^{\text{UHP}},
\fe
where
\ie
\begin{cases}
z'= F_I(w) = \left( \frac{1+iw}{1-iw} \right)^2 &\text{(unit disc)} \\[1pt]
z= f_I(w) = \frac{2w}{1-w^2} &\text{(UHP)}
\end{cases}
\fe
Equivalently, one can express the identity string field as $\bra{\mathbb{I}}=\bra{0}U_{f_I}$ where $U_{f_I}$ is the operator implementing the conformal transformation $z=f_I(w)$. 

The Ellwood invariant \cite{Ellwood:2008jh} can be defined as the 1-vertex with the extra insertion of an on-shell closed string  vertex operator $V(i,-i)$ (at the string midpoint) that is a primary of weight $(0,0)$. We denote it 
\be
W(\Psi, V) \equiv \vv{ \mathbb{I} | V(i,-i) | \Psi} = \vv{V(i,-i) ~ f_I \circ \Psi(w=0)}^{\text{UHP}}
\ee
Note that as the string field carries ghost number 1, the closed string vertex operator $V$ must carry ghost number 2 to give a nonzero result.

\section{Conservation laws for vertices}
\label{app:sympCons}

The conservation laws used for evaluating the symplectic form (\ref{OSFTsymp}) in level truncation are derived as follows. The correlators in question are of the form 
\ie
\vv{V(i,-i) ~ f_1 \circ A(w_1=0) ~ f_2 \circ B(w_2=0)}^{\text{UHP}}.
\fe
We take $z$ to be the coordinate on the UHP, and $w_1, \, w_2$ the local coordinates on the UHD containing the open string fields $A, B$ respectively. They are related by the coordinate transformations
\ie
\label{fiDef}
z=f_1(w_1) = \frac{1+w_1}{1-w_1} \qquad z=f_2(w_2)= - \frac{1-w_2}{1+w_2},
\fe
mapping the pair of $w_i$-UHDs to the right and left half of the $z$-UHP, respectively. Note that the origins of the $w_1$- and $w_2$-UHD are mapped to $z=1$ and $z=-1$. The $w_i$-semicircles are mapped to the imaginary $z$-axis.

In our application, $V$ is a conformal primary while $A$ and $B$ are not primaries in general. We use Ward identities to replace raising operators in $A$ or $B$ with linear combinations of lowering operators and constants acting on the remaining vertex operators, eventually reducing the correlator to that of primaries. We begin with a correlator of the form 
\ie
\vv{V(i,-i) ~ f_1 \circ (\phi_{-m}A)(w_1=0) ~ f_2 \circ B(w_2=0)}^{\text{UHP}},
\fe
where  the field $\phi$ is either $i\sqrt{2}\partial X^0$, $T^{(25)}$, $b$, or $c$.  We can express the mode $\phi_{-m}$ as
\ie\label{eq:phimode}
 \phi_{-m}= \int_C \left[ \frac{dw}{2 \pi i} \frac{\phi(w)}{w^{m-h+1}} -  \frac{d\bar{w}}{2 \pi i} \frac{\tilde{\phi}(\bar{w})}{\bar{w}^{m-h+1}} \right],
\fe
where $h$ is the weight of $\phi$ and $C$ is a CCW semicircular contour enclosing $w_1=0$. Deforming the contour $C$ away to infinity in the UHP, we must take into account the coordinate transformation between the $w_1$-chart and the $z$-UHP, as well as the residue contributions when the contour moves past the $V$ and $B$ insertions.

Using $\frac{dz}{dw_1}=\frac{2}{(1-w_1)^2}=\frac{(z+1)^2}{2}$, (\ref{eq:phimode}) is written in the $z$-coordinate as
\ie
\int_{C_1} \frac{dz}{2 \pi i} \frac{1}{2^{h-1}} (z+1)^{m+h-1} (z-1)^{-m+h-1} \vv{V(i,-i) \phi(z) ~ f_1 \circ A ~ f_2 \circ B}^{\text{UHP}}
+ (\text{anti-holomorphic}),
\fe
where $C_1$ is now a CCW semicircle in the UHP enclosing $z=1$. The residue contribution from $\phi$ colliding with $V$ can be determined from the OPE
\ie
\phi(z) V(i,-i) &\sim \sum_k \frac{C_k V_k(i,-i)}{(z-i)^{h+h_V-h_k}},\\
\tilde\phi(\bz) V(i,-i) &\sim \sum_{\bar{k}} \frac{\tilde{C}_{\bar{k}} \tilde{V}_{\bar{k}}(i,-i)}{(\bz+i)^{h+h_V-h_{\bar{k}}}},
\fe
where $h_V, \, h_k,\, h_{\bar{k}}$ are the weights of $V, \, V_k,\, V_{\bar{k}}$ respectively. The residue contribution from $\phi$ colliding with $B$ is a correlator involving $\phi_m B$. The result is
\ie
&\vv{V(i,-i) ~ f_1 \circ (\phi_{-m}A)(w_1=0) ~ f_2 \circ B(w_2=0)}^{\text{UHP}}  \\
& = (-1)^{m+h}\vv{V(i,-i) ~f_1 \circ A(w_1=0) ~ f_2 \circ (\phi_m B)(w_2=0) }^{\text{UHP}} \\
&~~~  -\sum_k  \underset{z\to i}{\text{Res}} \left( \frac{(z+1)^{m+h-1}(z-1)^{-m+h-1}}{2^{h-1}(z-i)^{h+h_V-h_k}} \right) C_k \vv{V_k(i,-i) ~f_1 \circ A~ f_2 \circ B }^{\text{UHP}}\\
&~~~ - \sum_{\bar{k}}  \underset{\bz\to -i}{\text{Res}} \left( \frac{(\bz+1)^{m+h-1}(\bz-1)^{-m+h-1}}{2^{h-1}(\bz+i)^{h+h_V-h_{\bar{k}} }} \right) \tilde{C}_{\bar{k}} \vv{\tilde{V}_{\bar{k}}(i,-i) ~f_1 \circ A~ f_2 \circ B }^{\text{UHP}}.
\fe
Specializing to the case $V = D = {1\over V_{X^0}} \left(c \partial X^0 + \tilde{c} \bar{\partial} X^0 \right)$ of our interest, the replacement rules are
\ie
\label{alphaLaw}
\vv{D(i,-i) ~ f_1 \circ (\alpha_{-m}A) ~ f_2 \circ B}^{\text{UHP}} = & (-1)^{m+1}\vv{D(i,-i) ~f_1 \circ A ~ f_2 \circ (\alpha_m B)}^{\text{UHP}} \\
&-i \frac{(-i)^{m} }{\sqrt{2} } m  \vv{\frac{c(i)}{ V_{X^0}} ~f_1 \circ A~ f_2 \circ B }^{\text{UHP}}\\
&-i \frac{(i)^{m} }{\sqrt{2} } m \vv{ \frac{\tilde{c}(-i) }{ V_{X^0}} ~f_1 \circ A~ f_2 \circ B }^{\text{UHP}},
\fe
\ie
\label{cLaw}
\vv{D(i,-i) ~ f_1 \circ (c_{-m}A) ~ f_2 \circ B}^{\text{UHP}} = & (-1)^{m-1}\vv{D(i,-i) ~f_1 \circ A ~ f_2 \circ (c_m B)}^{\text{UHP}},
\fe
\ie
\label{bLaw}
\vv{D(i,-i) ~ f_1 \circ (b_{-m}A) ~ f_2 \circ B}^{\text{UHP}} = & (-1)^{m+2}\vv{D(i,-i) ~f_1 \circ A ~ f_2 \circ (b_m B)}^{\text{UHP}} \\
&+ (-i)^m  \vv{\frac{\partial X^0(i)}{ V_{X^0}} ~f_1 \circ A~ f_2 \circ B }^{\text{UHP}}\\
&+ (i)^m  \vv{\frac{\bar{\partial} X^0 (-i)}{ V_{X^0}} ~f_1 \circ A~ f_2 \circ B }^{\text{UHP}},
\fe
\ie
\label{L25Law}
\vv{D(i,-i) ~ f_1 \circ (L^{(25)}_{-m}A) ~ f_2 \circ B}^{\text{UHP}} = & (-1)^{m+2}\vv{D(i,-i) ~f_1 \circ A ~ f_2 \circ (L^{(25)}_m B)}^{\text{UHP}}.
\fe

\section{Numerical implementation of level truncation}
\label{app:computationalDetails}

Here we describe some details of the numerical implementation of the level truncation computations of section \ref{sec:energyperturb}. Our approach is similar to that of \cite{Kudrna:2019xnw} (section 3), with one difference being that we impose a cutoff on the total oscillator level rather than the total weight (cf. (\ref{eq:levelDef})).

The main hurdle is to calculate the cubic vertices for all triplets of basis states \ref{generalState}, the symplectic form correlators $\vv{D(i) ~ f_1 \circ A(w_1=0) ~ f_2 \circ B(w_2=0)}^{\text{UHP}}$ for $(A,B)$ all pairs of basis states, and the Ellwood invariant  $W(A,V)$, defined in \ref{ellwoodUHP} for any individual basis state $A$. The cubic vertices can then be used to calculate the perturbative solution $\Psi_i$ as described in \ref{sec:leveltrunc}. Once $\Psi_i$ are obtained, the symplectic form and Ellwood invariant for the perturbative string field solution can be calculated using the symplectic form correlators and Ellwood invariants of basis states. One convenience is that all three of the ingredients, cubic vertices, symplectic form correlators, and Ellwood invariants, decouple between the $c=25$ Liouville, timelike boson, and ghost sectors, so they can be separately calculated for all three, and combined as needed by multiplication. To implement this on a computer, it is convenient to define a canonical ordering for the basis states in each sector, in order of increasing level of oscillators. For example
\be
1 \to 1, \,\, L^{(25)}_{-2} \to 2, \,\, L^{(25)}_{-3} \to 3, \,\,L^{(25)}_{-4} \to 4, \,\, L^{(25)}_{-2}L^{(25)}_{-2} \to 5,  \,\, \text{etc.}
\ee
in order of increasing level. For the boson sector, where parity is an important consideration, we numbered the $\alpha_{-\un{M}}$ which had an even number of oscillators with increasing positive integers and the  $\alpha_{-\un{M}}$ which had an odd number of oscillators with decreasing negative integers.
\ie
1 \to 1, \,\, \alpha_{-1}\alpha_{-1} \to 2, \,\,\alpha_{-2}\alpha_{-1} \to 3, \,\, \alpha_{-3}\alpha_{-1} \to 4, \,\, \alpha_{-2}\alpha_{-2} \to 5,  \,\, \text{etc.}\\
\alpha_{-1} \to -1, \,\, \alpha_{-2} \to -2, \,\,\alpha_{-3} \to -3, \,\, \alpha_{-1}\alpha_{-1}\alpha_{-1} \to -4, \,\, \alpha_{-4} \to -5,  \,\, \text{etc.}
\fe
The algorithm proceeds as follows:

\noindent 1. Calculate the ingredients: cubic vertices, symplectic form correlators, and Ellwood invariants for all the necessary triplets, pairs, and single basis states. This can be donse separately in the Liouville, timelike boson, and ghost sectors. Use the conservation laws described in \cite{Kudrna:2019xnw} and \ref{app:sympCons} to reduce the calculation of an ingredient at a given level to a linear combination of the same ingredient at lower levels. It is worth noting that  for the cubic vertex of three basis states at levels $(L_{v1}, L_{v2}, L_{v3})$ the cubic vertices that appear when applying the conservation laws have levels $(L_{v1}' \leq L_{v1}, L_{v2}' \leq L_{v2}, L_{v3}' \leq L_{v3})$ while the total level must strictly decrease $L_{v1}' + L_{v2}' + L_{v3}' < L_{v1} + L_{v2} + L_{v3}$. Similarly for the symplectic form correlators, except with just two basis states. Because the Liouville, boson, and ghost sectors are calculated separately, the level $L_v$ here always refers to the level of oscillators \textit{in that particular sector only}.

One can set up a loop over all increasing values of $L_{v1}, L_{v2}, L_{v3}$ each from $0$ to ${\bf L}$. Because the recursive conservation laws will always decrease the total level and never raise the levels of the first, second, or third basis states individually, the recursion will never have to go deeper than one step if one calculates \textbf{all} the cubic vertices at each step of this loop. This is important when running a parallel version of the algorithm, because it reduces how often two threads will be repeating the same calculation.

In practice, we do not actually need to calculate all the cubic vertices for the boson sector at each level. In this case the recursive calls may go deeper than one step, but this is not an issue. The boson sector also carries the additional complication of having an $f_i$ associated with each basis state. This will be treated below.

\noindent 2. Calculate the coefficients $A_{i,v,f}$ appearing in $\Psi_i$ using the cubic vertices. This does not require any recursive calls, so one does not have to be careful about the order in which they are calculated. The only requirement is that $\Psi_2$ be calculated before $\Psi_3$ before $\Psi_4$ etc. In the parallel implementation, we set up a loop over increasing $i = 2,\, 3,\, 4$. Within each run of the loop, the task of calculating the $A_{i,v,f}$ was split up evenly between all the threads since for a fixed $i$, no $A_{i,v,f}$ needs the value of any other.

\noindent 3. Calculate the symplectic form or Ellwood invariant using the symplectic form correlators or Ellwood invariants for basis states from step 1 and the coefficients $A_{i,v,f}$ from step 2. Like in step 2, all the recursive work has been done, and when implementing this in parallel, it is easiest to divide the work evenly among all the threads without worrying about the order.

We implemented this algorithm using Mathematica and C++, using the WSTP library to communicate between the two.

As the level cutoff ${\bf L}$ is increased, the number of basis states grows, and the number of cubic vertices grows quite quickly. This is the main reason why a parallelized implementation of the algorithm is necessary. 
The approximate storage costs at each level are

\begin{center}
\begin{tabular}{| m{4 cm} || c c c c c c c|}
\hline
 Level $\bf L$  & 8 & 10 & 12 & 14 & 16 & 18 &20 \\ 
 \hline
 Cost & 5 MB & 40 MB & 288 MB & 1.8 GB & 10 GB & 54 GB & 261 GB  \\  
 \hline
 \end{tabular}
\end{center}
Where the most space-consuming part are the timelike boson cubic vertices (where the basis states carry an extra label $f$). There is also the memory cost of running one Wolfram kernel on each thread, which is not estimated here. One node on the Harvard FASRC cluster has 184 GB RAM, so level 18 is the limit without writing and reading from hard disk/SSD memory or using multiple nodes together.

 We ran our program on a single node of the Harvard FASRC cluster, which has 184 GB RAM and 48 processor cores. Running on the full 48 cores, the runs took
 \begin{center}
\begin{tabular}{| m{4 cm} || c c c c c|}
\hline
 Level $\bf L$  & 8 & 10 & 12 & 14 & 16  \\ 
 \hline
Time & 20 sec & 4.5 min & 19 min & 2.5 hr & 21.5 hr   \\  

 \hline
 \end{tabular}
\end{center}
to execute. A simple experiment at level 12 shows that the speedup due to multiple cores is not exactly linear, but is not far off.
 \begin{center}
\begin{tabular}{| m{5 cm} || c c c c c|}
\hline
 Number of CPUs  & 48 & 24 & 12 & 6 & 3  \\ 
 \hline
Time to run level 12 & 19 min & 34 min & 64 min & 120 min & 237 min   \\  
 \hline
 \end{tabular}
\end{center}
These data fit the curve $\text{Time} = (625.15\text{ min})(\text{CPUs})^{-0.91}$ so the program runs 1.88 times faster when the number of CPUs is doubled.
This is to be expected, since the recursive part of the algorithm involves threads waiting for each other, and there is also an associated overhead time cost to managing multiple threads.
\bibliographystyle{JHEP}
\bibliography{symplectic}

\end{document}